\documentclass{acm_proc_article-sp}
\usepackage{bibentry}
\usepackage{amssymb}
\usepackage{balance}  
\usepackage{times}
\usepackage{algorithm}% 
\usepackage{algpseudocode}%
\usepackage{mathptmx}
\usepackage{float}
\usepackage[caption = false]{subfig}
\usepackage{graphicx}
\newtheorem{example}{Example}
\newtheorem{definition}{Definition}
\newcommand{\topic}[1]{\vspace{4pt} \noindent \underline{\bf #1}}

\newcommand{\ceil}[1]{\left\lceil #1 \right\rceil}
\newcommand{\floor}[1]{\left\lfloor #1 \right\rfloor}
\newcommand*\mean[1]{\bar{#1}}

\begin{document}

\setlength{\pdfpageheight}{\paperheight}
\setlength{\pdfpagewidth}{\paperwidth}

%\conferenceinfo{Submission to SoCC '15}{August, 2015, Kohala Coast, Hawaii, USA} 
%\copyrightyear{2015} 
%\copyrightdata{978-1-nnnn-nnnn-n/yy/mm} 
%\doi{nnnnnnn.nnnnnnn}

\title{Historical Graph Data Management}
\title{Storing and Analyzing Historical Graph Data at Scale}

\numberofauthors{2} 
\author{
\alignauthor Udayan Khurana\\
       \affaddr{IBM TJ Watson Research Center}\\
%       \affaddr{Wallamaloo, New Zealand}\\
       \email{ukhurana@us.ibm.com}       
\alignauthor Amol Deshpande\\
       \affaddr{University of Maryland}\\
       \email{amol@cs.umd.edu}
}

\maketitle
\begin{abstract}
The work on large-scale graph analytics to date has largely focused on the study of static properties of graph snapshots.
However, a static view of interactions between entities is often an oversimplification of several complex phenomena 
like the \textit{spread of epidemics}, \textit{information diffusion}, \textit{formation of online communities}, and so on.
Being able to find temporal interaction patterns, visualize the evolution of graph properties, or even simply compare them 
across time, adds significant value in reasoning over graphs. 
However, because of lack of underlying data management support, an analyst today has to manually navigate the added temporal
complexity of dealing with large evolving graphs. 
%The key desiderata therein include: compact storage of large historical graph traces, efficient retrieval of graph primitives at query time, a declarative language or high-level interface to express complex temporal graph analytical tasks, and ability to execute those tasks in a scalable manner.
In this paper, we present a system, called \textit{Historical Graph Store}, that enables users to store large
volumes of historical graph data and to express and run complex temporal graph analytical tasks against that data.
It consists of two key components: a {\em Temporal Graph Index} (TGI), that compactly stores large volumes of 
historical graph evolution data in a partitioned and distributed fashion; it provides support for retrieving snapshots of the graph as 
of any timepoint in the past or evolution histories of individual nodes or neighborhoods; and a Spark-based {\em Temporal Graph Analysis Framework} (TAF), 
for expressing complex temporal analytical tasks and for executing them in an efficient and scalable manner. 
Our experiments demonstrate our system's efficient storage, retrieval and analytics across a wide variety of queries on large volumes of historical graph data.

\end{abstract}

\section{Introduction}
Graphs are useful in capturing behavior involving interactions between
entities. Several processes are naturally represented as graphs -- social
interactions between people, financial transactions, biological interactions
among proteins, geospatial proximity of infected livestock, and so on. Many
problems based on such graph models can be solved using well-studied algorithms
from graph theory or network science. Examples include finding driving routes
by computing shortest paths on a network of roads, finding user communities 
through dense subgraph identification in a social network, and many others.
Numerous graph data management systems have been
developed over the last decade, including specialized graph database systems
like Neo4j, Titan, etc., and large-scale graph processing frameworks such
%For instance, the relative importance of
%a webpage used by search engines, is found through the page rank amongst a
%network of all pages from the world wide web; driving routes are found by
%computing shortest paths on the network of roads, and so on. Much of the
%application of graphs centric analysis in the past has been restricted to
%static models, i.e., ones where there is no account of change in the network
%structure or attributes. Apart from a strong theoretical foundation in graph
%theory, there is a considerable support for such analysis in form of tools and
%systems. 
%Graph stores such as Neo4J, Titan, etc., provide storage support and
%libraries implementing graph algorithms. Large graph processing frameworks such
as Pregel~\cite{pregel}, Giraph, %\footnote{http://giraph.apache.org/},
GraphLab~\cite{distgraphlab}, GraphX~\cite{graphx}, GraphChi~\cite{kyrola2012graphchi}, etc. %potentially scale up
%to graphs with millions of nodes and this is an active area of systems research
%and development. 

However much of the work to date, especially on cloud-scale graph data management systems, focuses 
on managing and analyzing a single (typically, current) static snapshot of the data. In the real world, however,
interactions are a dynamic affair and any graph that abstracts a real-world process changes over time.
For instance, in online social media, the friendship network on Facebook or the ``follows'' network on Twitter 
change steadily over time, whereas the ``mentions'' or the ``retweet'' networks change much more rapidly. 
Dynamic cellular networks in biology, evolving citation networks in publications, dynamic financial 
transactional networks, are few other examples of such data. Lately, we have seen an increasing merit in dynamic modeling and analysis of 
network data to obtain crucial insights in several domains such as cancer prediction~\cite{taylor2009dynamic}, epidemiology~\cite{gross2006epidemic}, organizational sociology~\cite{gulati1999interorganizational}, molecular biology~\cite{eisenberg2000protein}, 
information spread on social networks~\cite{lerman2010information} amongst others.

In this work, our focus is on providing the ability to analyze and to reason over the entire history of 
the changes to a graph. There are many different types of analyses of interest. 
For example, an analyst may wish to study the evolution of well-studied static graph properties such
as centrality measures, density, conductance, etc., over time. Another approach is through the search and 
discovery of temporal patterns, where the events that constitute the pattern are spread out over time. 
%In many application domains, we need the additional ability to analyze and reason over the historical 
%information. Temporal analysis of graphs may be done through different approaches. For instance, studying the evolution 
%of well-defined static graph properties such as centrality, density, flow, etc., is one such approach. Another 
%approach is through the search and discovery of temporal patterns, i.e., set of events that capture a story 
%in a chronological order. 
Comparative analysis, such as juxtaposition of a statistic over time, or perhaps, computing 
aggregates such as \textit{max} or \textit{mean} over time, possibly gives another style of knowledge discovery 
into temporal graphs. Most of all, a primitive notion of just being able to access past states of the graphs 
and performing simple static graph analysis, empowers a data scientist with the capacity to 
perform analysis in arbitrary and unconventional patterns.

%In the real world, however, interactions are a dynamic affair. Hence, any graph that abstracts a real world process, changes over time. 
%With the increase in availability of temporally annotated datasets, there is a also a greater abundance of temporal graph datasets. 
%Such datasets present a rich opportunity to obtain historical insights into the nature of various graphs. They also provides an opportunity to infer additional knowledge based on temporal analysis of graphs. For instance, the outbreak of an epidemic has an inevitable temporal component. A study determining the outbreak patterns needs to consider the order, and in fact, the actual times at which certain events occurred. Similarly, the rise of an influential figure in an information network, detecting suspicious patterns of server requests in a network, require a legitimate temporal modeling of the underlying data. A flattened, time oblivious model is an oversimplification and fails to capture the aspects of graph evolution. 
%

Supporting such a diverse set of temporal analytics and querying over large volumes of historical graph data 
requires addressing several data management challenges. Specifically, there is a want of techniques for storing
the historical information in a compact manner, while allowing a user to retrieve graph snapshots as of any
time point in the past or the evolution history of a specific node or a specific neighborhood. Further the
data must be stored and queried in a distributed fashion to handle the increasing scale of the data.
We must also develop an expressive, high-level, easy-to-use programming framework that will allow users to 
specify complex temporal graph analysis tasks, while ensuring that the specified tasks can be executed efficiently 
in a data-parallel fashion across a cluster.

%The lack of a principal theoretical foundation for temporal graph theory is a drawback in an effort to perform a wide variety of efficient temporal graph analytics. More prohibitive, however, is the limitation of the current set of available tools and technologies for graph storage, retrieval and processing to perform even simple temporal analysis at a reasonable scale.
%We believe that the data management challenges in this domain can be abstracted as follows: first, the \textit{storage} of large time-evolving graphs in a compact manner; second, efficient \textit{retrieval} of versions, snapshots, and other primitives; third, an expressive interface to \textit{specify} complex temporal graph analysis objectives; and finally, to be able to \textit{execute} graph analytics in a scalable and efficient environment.

In this paper, we present a graph data management system, called {\em Historical Graph Store (HGS)}, that provides
an ecosystem for managing and analyzing large historical traces of graphs.  HGS consists of two key distinct components. 
First, the {\em Temporal Graph Index (TGI)}, is an index that compactly stores the entire history of a graph by appropriately
partitioning and encoding the differences over time (called {\em deltas}).
These deltas are organized to optimize the retrieval of several temporal graph primitives such as 
neighborhood versions, node histories, and graph snapshots. TGI is designed to use a distributed key-value store to store the 
partitioned deltas, and can thus leverage the scalability afforded by those systems (our implementation uses Apache Cassandra\footnote{https://cassandra.apache.org}
key-value store). TGI is a tunable index structure, and we investigate the impact of tuning the different parameters through
an extensive empirical evaluation.
TGI builds upon our prior work on DeltaGraph~\cite{icdepaper},
where the focus was on retrieving individual snapshots efficiently; we discuss the differences between the two in 
more detail in Section \ref{sec:tgi}.

The second component of HGS is a \textit{Temporal Graph Analysis Framework (TAF)}, which provides an expressive library to specify 
a wide range of temporal graph analysis tasks and to execute them at scale in a cluster environment. 
The library is based on a novel set of \textit{temporal graph operators} that enable a user to analyze the history of a graph
in a variety of manners. The execution engine itself is based on Apache Spark~\cite{zaharia2010spark}, a large-scale in-memory 
cluster computing framework.

%XXX: Explicitly call out the contributions in the intro beyond providing the system details.

\topic{Outline:} The rest of the paper is organized as follows. In Section~\ref{sec:related}, we survey the related work on graph data stores, temporal indexing, and other topics relevant to the scope of the paper. In Section~\ref{sec:overview}, we provide a sketch of the overall system, including key aspects of the underlying components. We then present the Temporal Graph Index and the Temporal Graph Analytics Framework in detail in Section~\ref{sec:tgi} and Section~\ref{sec:taf}, respectively. In Section~\ref{sec:experiments}, we provide an empirical evaluation of the various system components such as the graph retrieval, scalability of temporal analytics, etc. We conclude with a summary and a list of future directions in Section~\ref{sec:conclusion}.

\section{Related Work}
\label{sec:related}

In the recent years, there has been much work on graph storage and graph processing
systems and numerous systems have been designed to address various aspects of graph
data management. Some examples include Neo4J, AllegroGraph~\cite{aasman2006allegro},
Titan\footnote{http://thinkaurelius.github.io/titan/}, GBase~\cite{kang2011gbase},
Pregel~\cite{pregel}, Giraph, GraphChi~\cite{kyrola2012graphchi}, GraphX~\cite{graphx},
GraphLab~\cite{distgraphlab}, and Trinity~\cite{shao2013trinity}. These systems use a variety
of different models for representation, storage, and querying, and there is a lack of
standardized or widely accepted models for the same.
Most graph querying happens through programmatic access to graphs in languages such as Java, 
Python or C++. Graph libraries such as Blueprints\footnote{https://github.com/tinkerpop/blueprints/wiki} 
provide a rich set of implementations for graph theoretic algorithms. SPARQL~\cite{perez2006semantics} is 
a language used to search patterns in linked data. It works on an underlying RDF representation of graphs. 
T-SPARQL~\cite{grandi2010t} is a temporal extension of SPARQL. He et al.~\cite{he:sigmod08}, provide 
a language for finding sub-graph patterns using a graph as a query primitive. 
Gremlin\footnote{https://github.com/tinkerpop/gremlin} is a graph traversal language over the property graph
data model, and has been adopted by several open-source systems.
For large-scale graph analysis, perhaps the most popular framework is the vertex-centric programming
framework, adopted by Giraph, GraphLab, GraphX, and several other systems; there have also been
several proposals for richer and more expressive programming frameworks in recent years.
%Each one of them address certain
%challenges in graph data management, using different models of storage and processing. Some of them
%focus on efficient disk storage, versus the others that are focused towards a large scale out.
%However, a standardized or widely accepted model of data representation or computation is still an
%open question. 
However, most of these prior systems largely focus on analyzing a single snapshot of the graph data,
with very little support for handling dynamic graphs, if any.

A few recent papers address the issues of storage and retrieval in dynamic graphs. In our prior
work, we proposed DeltaGraph~\cite{icdepaper}, an index data structure that compactly stores the history of all changes 
in a dynamic graph and provides efficient snapshot reconstruction. G*~\cite{gstar} stores multiple snapshots compactly by utilizing commonalities. Chronos~\cite{hant2014chronos,immortalgraph} is an in-memory system for processing dynamic graphs, with objective of shared storage and computation for overlapping snapshots. Ghrab et al.~\cite{ghrab2013analytics} provide a system of network analytics through labeling graph components.
Gedik et al.~\cite{6702469}, describe a block-oriented and cache-enabled system to exploit spatio-temporal locality for solving temporal neighborhood queries. 
Koloniari et al. also utilize caching to fetch selective portions of temporal graphs they refer to as partial views~\cite{koloniari2013partial}.
LLAMA~\cite{llama} uses multiversioned arrays to represent a mutating graph, but their focus is primarily on in-memory representation.
There is also recent work on streaming analytics over dynamic graph data~\cite{kineograph,graphinc}, but it typically 
focuses on analyzing only the recent activity in the network (typically over a sliding window).
Our work in this paper focuses on techniques for a wide variety of temporal graph retrieval and analysis on entire graph histories.

Temporal graph analytics is an area of growing interest. Evolution of shortest paths in dynamic
graphs has been studies by Huo et al.~\cite{huo2014efficient}, Ren et al.~\cite{RenEvolvGraph11},
and Xuan et al.~\cite{xuan2003computing}. Evolution of community structures in graphs has been of
interest as
well~\cite{asur2009event,berger2006framework,greene2010tracking,tang2008community}.
Change in page rank with evolving graphs~\cite{desikan2005incremental,bahmani2010fast}, and
the study of change in centrality of vertices, path lengths of vertex pairs,
etc.~\cite{pan2011path}, also lie under the larger umbrella of temporal graph analysis. Ahn et
al.~\cite{ahn2014task} provide a taxonomy of analytical tasks over evolving
graphs. Barrat et al.~\cite{barrat2008dynamical}, provide a good reference for studying several
dynamic processes modeled over graphs. %Kolaczyk's book on statistical analysis of
%graphs~\cite{kolaczyk2009statistical}, serves as a good reference for techniques and applications
%for graph analysis in general.
Our system significantly reduces the effort involved in building and deploying such analytics over large volumes of graph data.

% XXXX HERE

Temporal data management for relational databases was a topic of active research in the 80s and early 90s.
Snapshot index~\cite{Tsotras1995} is an I/O optimal solution to the problem of snapshot retrieval 
for transaction-time databases.  Salzberg and Tsotras~\cite{Salzberg1999} present a comprehensive
survey of temporal data indexing techinques, and discuss two extreme approaches 
to supporting snapshot retrieval queries, referred to as the \textit{Copy} and \textit{Log} approaches. While the copy approach relies on storing new copies of a snapshot upon every point of change in the database, the log approach relies on storing everything through changes. Their hybrid is often referred to as the \textit{Copy+Log} approach. 
%The paper presents a comprehensive survey of temporal data indexing techniques~\cite{Salzberg1999}.
%The temporal graph literature distinguishes between two different basic notions of time in databases --
%\textit{valid time} and \textit{transaction time}, which are considered orthogonal to each other. Valid time denotes the time
%period during which a fact is true with respect to the real world. Transaction time is the time
%when a fact is stored in the database. Database systems can be specific to either of the two times. There are also systems that incorporate both and are called bi-temporal databases. Another concept of \textit{application time} refers to an extraneous time quantity recorded by the database. 
We omit a detailed discussion of the work on temporal databases, and refer the interested reader to a representative set of references~\cite{Bolour92,DBLP:conf/sigmod/SnodgrassA85,Ozsoyoglu1995,Tansel1993,date2002temporal,tsql2,Salzberg1999}. 
Other data structures, such as Interval Trees~\cite{Arge1996} and Segment trees~\cite{Blankenagel1994} can also be used for storing temporal information. Temporal aggregation in scientific array databases~\cite{soroush2013time} is another related topic of interest, but the challenges there are significantly different.

%Miscellaneous systems and data management topics touch the scope of the work presented in this paper. For the reasons of space, we avoid a detailed survey of those topics. Some of the more relevant references, however are, data placement and co-location~\cite{kumar2013data}, and graph partitioning~\cite{shaopage}, ~\cite{miettinen2006using}.

\section{Overview}
\label{sec:overview}
%In this section, we discuss the key aspects of historical graph data management, followed by an overview of our proposed system, Historical Graph Store.
In this section, we introduce key aspects related to HGS. We begin with the data model, followed by the key challenges and concluding with an overview of the system.

\subsection{Data Model}
Under a discreet notion of time, a time-evolving graph
$G^T=(V^T,E^T)$ may be expressed as a collection of graph \textit{snapshots}
over different time points, $\{G^0 = (V^0, E^0), G^1, \dots, G^t \}$. The vertex set $V^i$ 
for a snapshot consists of a set of vertices (nodes), each of which has a unique identifier, 
and an arbitrary number of key-value attribute pairs. The edge sets $E^i$ consist of edges that each contain
references to two valid nodes in the corresponding vertex set $V^i$, information about the direction of the edge, and
an arbitrary list of key-value attribute pairs. A temporal graph can also be equivalently
described by a set of changes to the graph over time. We call an atomic change at a specific timepoint
in the graph an \textit{event}. The changes could be structural, such as the addition or the
deletion of nodes or edges, or be related to attributes such
as an addition or a deletion or a change in the value of a node or an edge
attribute. These approaches as well as certain hybrids have been used in the
past for the physical and logical modeling of temporal data. Our approach to
temporal processing in this paper is best described using a
\textit{node-centric} logical model, i.e., the historical graph is seen as a
collection of evolving vertices over time; the edges are considered as
attributes of the nodes.
 
 \begin{figure}
\includegraphics[width=\linewidth, trim=0 0 0 0]{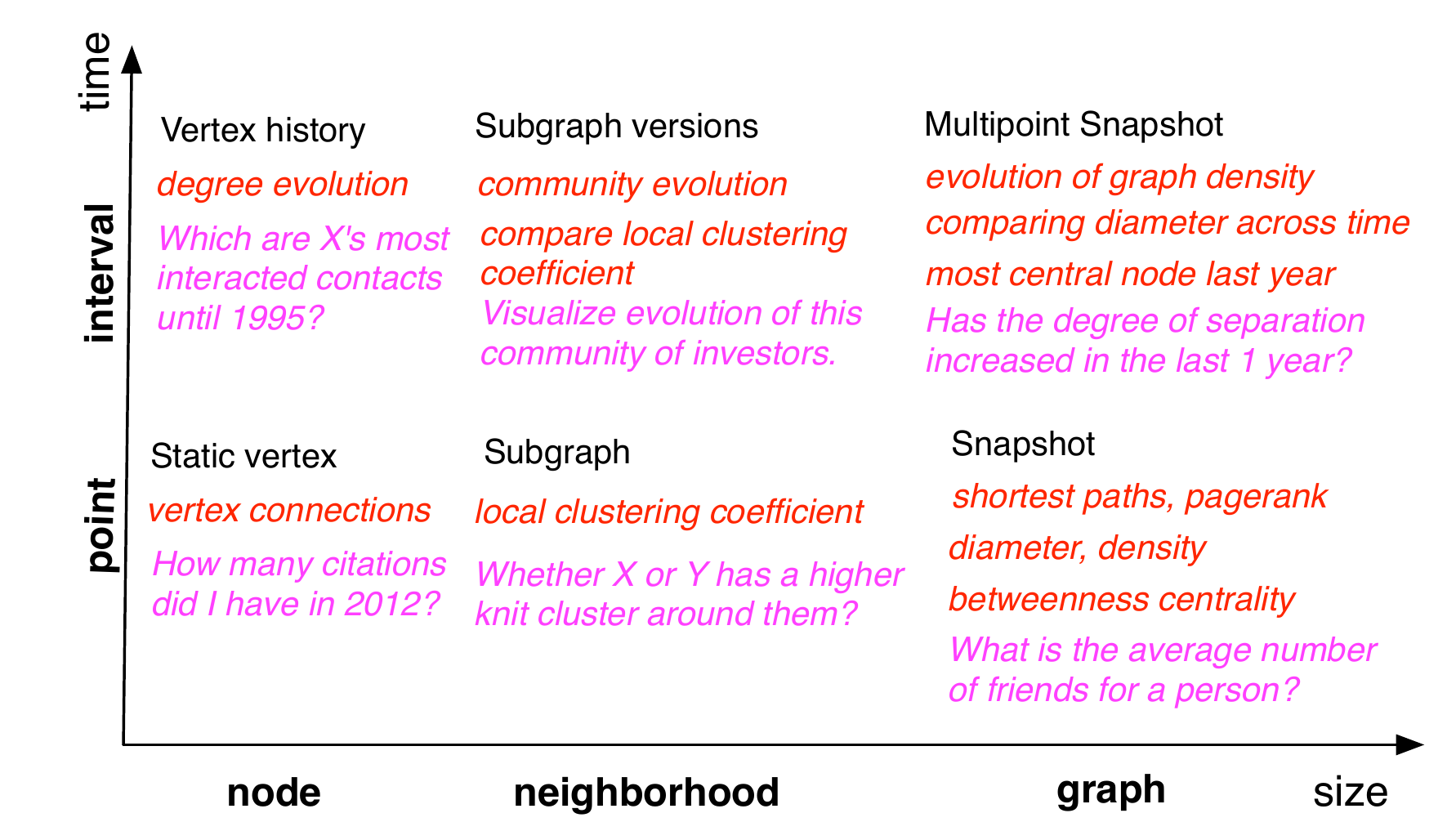}
\caption{The scope of temporal graph analytics can be represented across two different dimensions - time
and entity. The chart lists retrieval tasks (black), graph operations (red), example queries
(magenta) at different granularities of time and entity size.}
\label{fig:entity-time}
%\vspace{-10pt}
\end{figure}

\subsection{Challenges}
The nature of data management tasks in historical graph analytics can be categorized based on the scope of analysis using the dual 
dimensions of \textit{time} and \textit{entity} as illustrated with examples in Figure~\ref{fig:entity-time}. The temporal scope of an analysis task can range from a single point in time to a long interval; the entity scope can range from a single node to the entire graph.
While the diversity of analytical tasks provides a potential for a rich set of insights from historical graphs, it also 
poses several challenges in constructing a system that can perform those tasks. To the best of our knowledge, none of the existing systems address a majority of those challenges that
are described below:

 \topic{Compact storage with fast access:}  An natural tradeoff between index size and access latencies can be seen in the Log and Copy approaches for snapshot retrieval. 
Log requires minimal information for encoding the graph's history, but incurs large reconstruction costs. Copy, on the other hand, provides direct access, but at the cost of excessive storage. The desirable index should consume space of the order of Log index but provide near direct access like Copy.

 \topic{Time-centric versus entity-centric indexing:} For {\em point} access such as past snapshot retrieval, a time-centric indexing such as DeltaGraph or Copy+Log is suitable.
However, for version retrieval tasks such as retrieving a {\em node's history}, entity-centric indexing is the correct choice. Neither of the indexing approaches, however, are feasible in the opposite scenarios. Given the diversity of access needs, we require an index that works well with both styles of lookup at the same time.

 \topic{Optimal granularity of storage for different queries:} Query latencies for a graph also depends on the size of chunks in which the data is indexed. While larger granularities of storage incur wasteful data read for ``node retrieval'', a finely chunked graph storage would mean higher number of lookups and aggregation for a 2-hop neighborhood lookup. 
The physical and logical arrangement of data should take care of access needs of queries of all granularities.

 \topic{Coping with changing topology in a dynamic graph:} It is evident that graph partitioning is inevitable in the storage and processing of large graphs. 
However, finding the appropriate strategy to maintain workable partitioning on a constantly {\em changing} graph is another challenge while designing a historical graph index.

 \topic{Systematically expressing temporal graph analytics:} A platform for expressing a wide variety of historical graph analytics requires an appropriate amalgam of temporal logic and graph theory. Additionally, utilizing a vast body of existing tools in network science is an engineering challenge and opportunity.

 \topic{Appropriate abstractions for distributed, scalable analytics:} Parallelization is the key to scale up analytics for large network datasets. It is essential that the underlying data-representations and operators in the analytical platform be designed for parallel computing.

 \subsection{System Overview}
\begin{figure}
\includegraphics[width=\linewidth, trim=0 5 0 10]{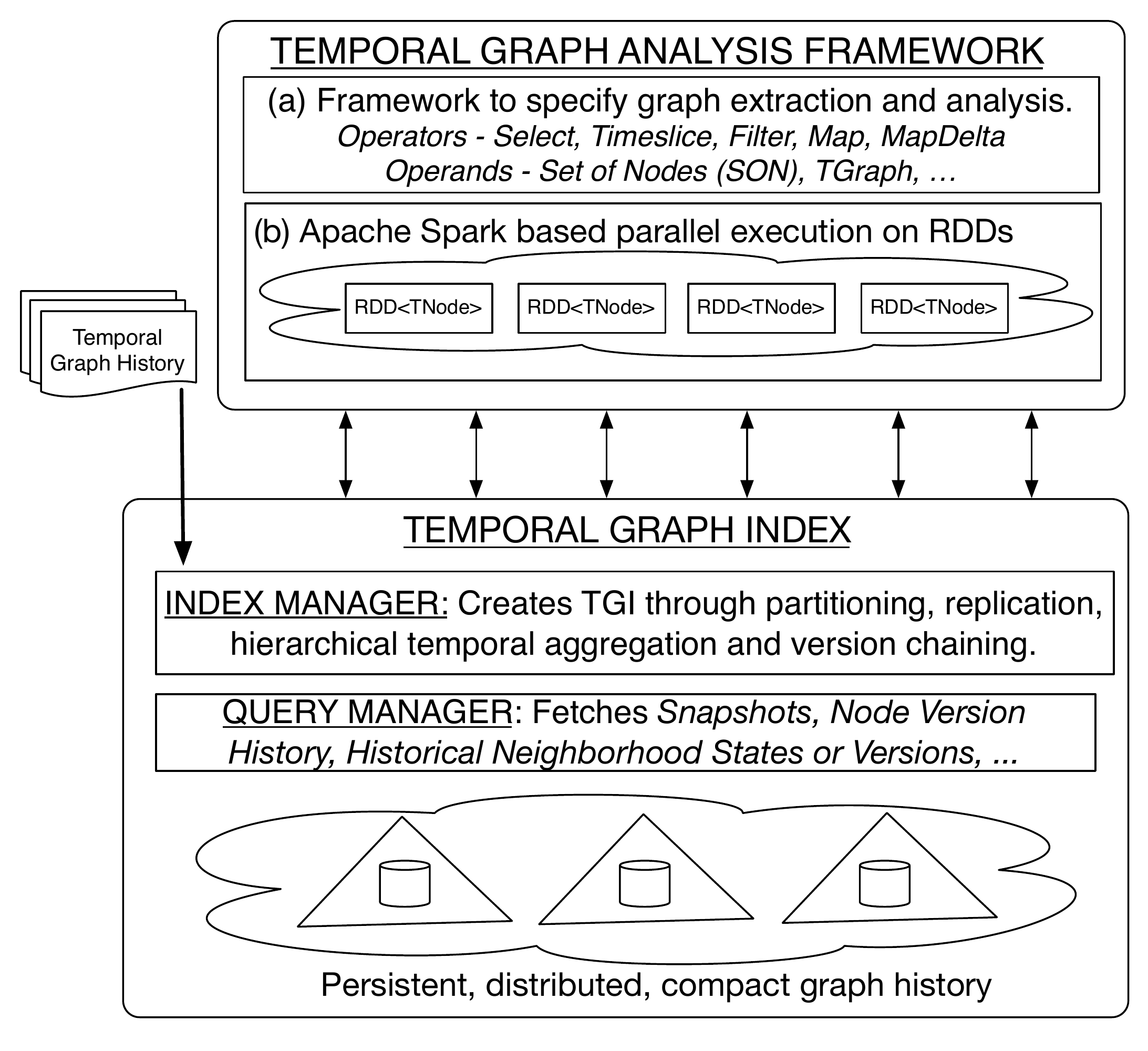}
\caption{System Overview}
\label{fig:overview}
%\vspace{-15pt}
\end{figure}

Figure~\ref{fig:overview} shows the architecture of our proposed Historical Graph Store. It consists of two main components:

\vspace{2pt}
\topic{Temporal Graph Index (TGI)} records the entire history of a graph compactly while enabling efficient retrieval of several temporal graph primitives. It encodes various forms of differences (called \textit{deltas}) in the graph, such as atomic events, changes in subgraphs over intervals of time, etc.  It uses specific choices of graph partitioning, data replication, temporal compression and data placement to optimize the graph retrieval performance. TGI uses the Apache Cassandra distributed key-value store as the backend to store the deltas.
%It extends and redesigns the notion of DeltaGraphs to to a broader spectrum of queries and in a much more scalable way. 
In Section~\ref{sec:tgi}, we describe the design details of TGI and the access algorithms. 
%We also analyze its performance and compare it to the alternatives. %others through a delta arithmetic, using two metrics that capture the notion of combined delta sizes and number of deltas that are fetched for a specific graph primitive type, respectively.

\vspace{2pt}
\topic{Temporal Graph Analytics Framework (TAF)} provides a \textit{temporal node}-centric abstraction for specifying and executing complex temporal network analysis tasks. We provide a Java and Python based library to specify the retrieval, computation and analysis on a \textit{set of (temporal) nodes (SoN)}. %An SoN can abstract a static or temporal graph or a subgraph. % , static, temporal, or a collection of them. 
Computational scalability is achieved by distributing tasks by node and time. TAF is built on top of Apache Spark for supporting scalable, in-memory, cluster computation and provides an option to utilize GraphX for static graph computation. In Section~\ref{sec:taf}, we describe the details of the library, query processing, parallel data fetch aspects of the system, along with a few examples of analytics.

\section{Temporal Graph Index}
\label{sec:tgi}
%Perhaps the simplest and the most compact way to store the history of a time
%evolving graph is to record all the changes that happened to the graph in a
%chronological order (this approach is often called \textit{Log}). However, it
%is fairly inefficient to retrieve either snapshots or versions of specific
%nodes on such a representation, because it requires reading a large volume of 
%information even if the result is small.
%More efficient indexes on time
%evolving datasets can be based along either of the two principle dimensions --
%time or entity. Time based indexing methods such as \textit{DeltaGraph} or
%\textit{Copy+Log}, provide fast access to the graph at specific time points,
%i.e., snapshots. On the other hand, entity-based indexing approaches, e.g., a
%\textit{node-centric temporal index}, would provide direct access to different
%versions of a specific node, but are fairly inefficient at fetching snapshots
%or subgraphs.

In this section, we investigate the issue of indexing temporal graphs. First, we  
introduce a \textit{delta framework}\footnote{A delta formalism provided by Ghandeharizadeh et
al.~\cite{Ghandeharizadeh:1996:HED:232753.232801} is an interesting related read on this topic.} to define any temporal index as
a set of different changes or \textit{deltas}. Using this framework, we are able to qualitatively compare the access costs and sizes of different alternatives for temporal graph indexing, including our proposed approach. 
We then present the Temporal
Graph Index (TGI), that stores the entire history of a large evolving network
in the cloud, and facilitates efficient parallel reconstruction for different
graph primitives. % such as past snapshots, neighborhood versions, etc., at runtime. 
TGI is a generalization of both entity and time-based indexing
approaches and can be tuned to suit specific workload needs. 
We claim that TGI is the minimal index that provides efficient access to a variety of primitives on a historical graph, ranging from past snapshots to versions of a node or neighborhood. We also describe the key partitioning strategies instrumental in scaling to large datasets across a cloud storage.
%We also describe
%the various access methods for graph primitives and compare the costs with
%alternatives using the delta arithmetic. 

\subsection{Preliminaries}
We start with a few preliminary definitions that help us formalize the notion of the delta
framework.
%\vspace{-5pt}
\begin{definition}[Static node] A \textit{static node} refers to the state of a vertex in a network at a specific time, and is defined
as a \textit{set} containing: (a) \textit{node-id}, denoted \textit{I} (an integer), (b) an edge-list, denoted \textit{E} (captured as a set of node-ids), (c) attributes, denoted \textit{A}, a map of key-value pairs.
\end{definition}

A \textit{static edge} is defined analogously, and contains the node-ids for the two endpoints and the edge direction in addition to a map of key-value pairs. Finally,
a \textit{static graph component} refers to either a static edge or a static node.

%\begin{definition}[Static edge] A \textit{static edge} corresponds to the state of an edge at a specified point in time. It is a set of the following objects: (a) an ordered pair of node-ids, ($n_1$ and $n_2$), (b) an edge-id, $I$ which is derived from the two node-ids, $n_1$ and $n_2$, $I=\{n_1,n_2\}$, (c) direction of the edge, (c)  attributes, $A$, is a map of string key, value pairs.
%\end{definition}

%\begin{definition}[Static graph component] A \textit{static graph component} refers to either a static node or a static edge. A collection of graph components can refer to a mixed set of both.
%\end{definition}
%\vspace{-5pt}
\begin{definition} [Delta] A Delta ($\Delta$) refers to either: (a) a static graph component (including the empty set), or (b) a difference, sum, union or intersection of two deltas. 
\end{definition}
%\vspace{-10pt}
\begin{definition} [Cardinality and Size] The cardinality \\and the size of a $\Delta$ are the unique and total number of static node or edge descriptions within it, respectively.
\end{definition}
%\vspace{-10pt}

\begin{definition} [$\Delta$ Sum] A sum (+) over two deltas, $\Delta_1$ and
$\Delta_2$, i.e., $\Delta_s = \Delta_1 + \Delta_2$ is defined over graph
components in the two deltas as follows: (1) $\forall  gc_1 \in \Delta_1$, if
$\exists gc_2 \in \Delta_2\ s.t.\ gc_1.I = gc_2.I$, then we add $gc_2$ to
$\Delta_s$, (2) $\forall gc_1 \in \Delta_1\ s.t.\ \nexists gc_2 \in \Delta_2\ s.t.\ gc_1.I = gc_2.I$,
we add $gc_1$ to $\Delta_s$, and (3) analogously the components present only in $\Delta_2$ are
added to $\Delta_s$.
\end{definition}

%
%$gc_s.I = gc_2.I$ and $gc_s$ 
%
 %$ \exists gc_s \in \Delta_s$ such that  $gc_s.I = gc_2.I$, and contains components of $gc_1$ overriden by changes in $gc_2$. For all the graph components with no matching node-id in the other set, $\forall gc_1 \in \Delta_1$, if $\nexists gc_2 \in \Delta_2$, such that $gc_1.I = gc_2.I$, and vice-versa are included as it is in the resulting delta. 
 
Note that: $\Delta_1 + \Delta_2 = \Delta_2 +\Delta_1$ is not necessarily true due the order of changes. We also note that: $\Delta_1 + \emptyset = \Delta_1$, and $(\Delta_1 + \Delta_2) + \Delta_3 = \Delta_1 + (\Delta_2 + \Delta_3)$. Analogously, difference(-) is defined as a set difference over different components of the two deltas.  $\Delta_1 - \phi = \Delta_1$ and  $\Delta_1 - \Delta_1 = \phi$, are true, while, $\Delta_1 - \Delta_2 = \Delta_2 - \Delta_1$, does not necessarily hold.

%%\vspace{-5pt}
%\begin{definition} [$\Delta$ Difference] A difference(-) is defined as a set difference over different components of the two deltas. $\Delta_1 - \phi = \Delta_1$ and  $\Delta_1 - \Delta_1 = \phi$, are true, while, $\Delta_1 - \Delta_2 = \Delta_2 - \Delta_1$, does not necessarily hold.
%\end{definition}

\begin{definition} [$\Delta$ Intersection] An intersection of two $\Delta$s is defined as a set intersection over the the components of two deltas. $\Delta_1 \cap \phi = \phi$, is true for any delta. Similarly, union of two deltas $\Delta_{\cup} = \Delta_1 \cup \Delta_2$, consists of all elements from $\Delta_1$ and $\Delta_2$. The following is true for any delta: $\Delta_1 \cup \phi = \Delta_1$.

\end{definition}

%\begin{definition} [$\Delta$ Union] A union of two deltas $\Delta_{\cup} = \Delta_1 \cup \Delta_2$, consists of all elements from $\Delta_1$ and $\Delta_2$. The following is true for any delta: $\Delta_1 \cup \phi = \Delta_1$.
%\end{definition}
\vspace{0pt}
\noindent
Next we discuss and define some specific types of $\Delta$s:
%%\vspace{-5pt}
\begin{example}[Event] An \textit{event} is the smallest change that happens to a graph, i.e., addition or deletion of a node or an edge, or a change in an attribute value. An event is described around one time point. As a $\Delta$, an event  concerning a graph component $c$, at time point $t_e$, is defined as the difference of state of $c$ at and before $t_e$, i.e., $\Delta_{event}(c,t_e) =  c(t_e) - c(t_e-1)$.
\end{example}
%\vspace{-5pt}
\begin{example}[Eventlist] An \textit{eventlist} delta is a chronologically sorted set of event deltas. An eventlist's scope may be defined by the time duration, $(t_s,t_e]$, during which it defines all the changes that happened to the graph.
\end{example}
%\vspace{-5pt}
\begin{example}[Partitioned Eventlist] An \textit{partitioned eventlist} delta is an eventlist constrained by the scope of a set of nodes (say a set of nodes, $\mathcal N = \{N_1, N_2, \dots\})$ apart from the time range constraint $(t_s,t_e]$.
\end{example}
%\vspace{-5pt}
\begin{example} [Snapshot] A snapshot, $\mathcal G^{t_a}$ is the state of a graph $\mathcal G$ at a time point $t_a$. As a $\Delta$, it is defined as the difference of the state of the graph at $t_a$ from an empty set, $\Delta_{snapshot}({\mathcal G}, t_a)= G(t_a)-G(-\infty)$. 
\end{example}
%\vspace{-5pt}
\begin{example} [Partitioned Snapshot] A partitioned snapshot is a subset of a snapshot. It is identified by a subset of all nodes, $\mathcal P$ in graph, $\mathcal G$ at time, $t_a$. It consists of the state of all nodes at time $t_a$ and all the edges whose at least one of the end points lies in $\mathcal P$ at time, $t_a$. 
\end{example}

\subsection{Prior Techniques}

The prior techniques for temporal graph indexing use changes or differences in various forms to 
encode time-evolving datasets. We can express them in the $\Delta$ framework as follows. The {\bf \em Log} index is equivalent to a set of all {\em event} deltas (equivalently, a single {\em eventlist} delta encompassing the entire history). The {\bf \em Copy+Log} index can be represented as combination of: (a) a finite number of distinct {\em snapshot} deltas, and (b) {\em eventlist} deltas to capture
the change between successive snapshots.
Although we are not aware of a specific proposal for a {\bf \em vertex-centric} index, however, a natural approach would be to maintain a set of 
{\em partitioned eventlist} deltas, one for each node (with edge information replicated with the endpoints).
%The node-centric\footnote{We are not aware of a node-centric index in temporal graph literature. However, due to its distinct properties, it is worth the discussion on this topic.} eventlist approach can be described as a separate set of event deltas for each of the nodes.
The {\bf \em DeltaGraph} index, proposed in our prior work, is a tunable index with several parameters. For a typical setting of parameters, it can be seen as
equivalent to taking a Copy+Log index, and replacing the {\em snapshot} deltas in it with another set of deltas constructed hierarchically as follows: 
for every $k$ successive {\em snapshot} deltas, replace them with a single delta that is the intersection of those deltas and a set of difference deltas from the intersection to the original snapshots, and recursively apply this till you are left with a single delta.

%expressed as a set of (a) all event deltas, (b) a hierarchical structure over a finite number of snapshot deltas, using an intersection function on 2 ($k$ in general) at a time.

Table~\ref{tab:deltacompare} estimates the cost of fetching different graph primitives as the number and the cumulative size
of deltas that need to be fetched for the different indexes. The first column shows an estimate of the total storage space, which varies
considerably across the techniques.

%We now compare different indexing approaches using a $\Delta$ cost model that estimates the cost of fetching different graph primitives based on the number and the cumulative size of deltas that need to be fetched in Table~\ref{tab:deltacompare}.

\begin{table*}[htp]
\begin{center}
{\small
\begin{tabular}{|c|c|c|c|c|c|c|c|c|c|c|c|}
\hline
 &Index &\multicolumn{2}{|c|}{Snapshot} & \multicolumn{2}{|c|}{Static Vertex} & \multicolumn{2}{|c|}{Vertex versions} & \multicolumn{2}{|c|}{1-hop} & \multicolumn{2}{|c|}{1-hop Versions} \\ 
\cline{3-12}
 &Size & $\sum_{\Delta}{|\Delta|}$ & $\sum_{\Delta}{1} $  & $\sum_{\Delta}{|\Delta|}$ & $\sum_{\Delta}{1} $  & $\sum_{\Delta}{|\Delta|}$ & $\sum_{\Delta}{1} $  & $\sum_{\Delta}{|\Delta|}$ & $\sum_{\Delta}{1} $ & $\sum_{\Delta}{|\Delta|}$ & $\sum_{\Delta}{1} $   \\
\hline
\hline
 Log & $|G|$ & $|G|$ & $|G| \over |E|$  & $|G|$ & $|G| \over |E|$ & $|G|$ & $|G| \over |E|$ & $|G|$ & $|G| \over |E|$ & $|G|$ & $|G| \over |E|$  \\
\hline
Copy & $|G|^2$ & $|S|$ & $1$ & $|S|$ & $1$ & $|S||G|$ & $|G|$ & $|S|$ & $1$ & $|S||G|$ & $|G|$\\ 
\hline
Copy+Log & ${|G|^2 \over |E|}$ & $|S|+|E|$ & $2$  & $|S|+|E|$ & 2 & $|G|$ & $|G| \over |E|$ &  $|S|+|E|$ & 2& $|G|$ & $|G| \over |E|$\\
\hline
Node Centric & $2|G|$ & $2.|G|$ & $|N|$ & $|C|$ & $1$ & $|C|$ & 1 &  $|R|.|V|$ & $|R|$ & $|R|.|V|$ & $|R|$\\
\hline
DeltaGraph &$X_1^*$& h.$|S|+|E|$ & $2h$  & $h.|S|+|E|$ & $2h$ & $|G|$ & $|G| \over |E|$ &  $h.(|S|+|E|)$ & 2h & $|G|$ & $|G| \over |E|$\\
\hline
TGI &$X_2^{**}$& h.$|S|+|E|$ & $2h$  & ${h.|S| \over p}+{|E| \over p}$ & $2h$ & $|V|(1+{|S| \over p})$ & $|V|+1$ &  ${h.(|S|+|E|) \over p}$ & 2h & $|V|(1+{|S| \over p})$ & $|V|+1$\\
\hline
\end{tabular}
}
\end{center}
\caption{Comparison of access costs for different retrieval queries and index storage on various temporal indexes. $|G|=$number of changes in the graph; $|S|=$size of a snapshot; $h=$ height and $|E|=$ eventlist size in C+L, DG or TGI; $|V|=$number of changes to a node; $|R|$=numbers of neighbors of a node; $p$= number of partitions in TGI. The metrics used are the sum of delta cardinalities ($\sum_{\Delta}{|\Delta|}$) and the number of deltas ($\sum_{\Delta}{1}$). For reasons of space, $^*X_1=|G|(h+1)$;  $^{**}X_2=|G|(2h+3)$.}
\label{tab:deltacompare}
\end{table*}%

\subsection{Temporal Graph Index: Definition}
Given the above formalism, a Temporal Graph Index for a graph $\mathcal G$ over a time period $T=[0, t_c]$ is described by 
a collection of different $\Delta$s as follows: 
\begin{itemize}
\item[(a)] Partitioned Eventlists: A set of partitioned eventlist $\Delta$s, $\{E_{tp}\}$, 
where $E_{tp}$ captures the changes during the time interval $t$ belonging to partition $p$.
%describing all the changes that happened in the graph. A partitioned eventlist describes the changes during the time interval $t$, belonging to partition $p$.

\item[(b)] Derived Partitioned Snapshots: Consider $r$ distinct time points, $t_i$, where $1 \le i \le r$, $t_i \in T$, %and $t_1 = 0$, and $t_i \ne t_k, \forall i \ne k$. 
For each $t_i$, we consider $l$ partition $\Delta$s, $P_{j}^i$, $1< j <l$, such that $\cup_{j} P_{j}^i = {\mathcal G^{t_i}}$. There exists a function that maps any node-id(I) in $\mathcal G^{t_i}$ to a unique partition-id($P_j^i$), $f_i: I \rightarrow P_j^{i}$. 
With a collection of $P_j^{i}$ over $T$ as leaf nodes, we construct a hierarchical tree structure where a parent is the intersection of children deltas. The difference of each parent from its child delta is called as a {\em derived partitioned snapshot} and is explicitly stored. Note that $P_j^{i}$'s are not explicitly stored. This is the same as DeltaGraph, with the exception of partitioning.
%These $P_j^{i}$, called {\em partitioned snapshots} are not explicitly stored. Instead, the {\em derived partitioned snapshots}, constructed in the following way are stored. Given a set of partitioned snapshots, we create a hierarchical structure similar to the DeltaGraph. Based on the partitioned snapshots, create parent snapshots per partition, using an \textit{intersection} function. Further, we compute the difference between a snapshot and its parent, and call it the derived partitioned snapshot deltas. These are the deltas that we actually store in the index.

\item[(c)] Version Chain: For all nodes $\mathcal N$ in the graph $\mathcal G$, we maintain a chronologically sorted list of pointers to all the references for that node in the delta sets described above (a and b). For a node $I$, this is called a \textit{version chain}($VC_I$). 
%There may be \textit{auxiliary version chains}($AVC_I^{p}$) that reflect change in a set of properties $p$ of the node $I$.
\end{itemize}

%\end{definition}

\begin{figure}
\centering
\subfloat[TGI deltas - partitioned eventlists, snapshots and derived snapshots. The (dotted) bounded deltas are not stored. ]{\includegraphics[width = .50\textwidth]{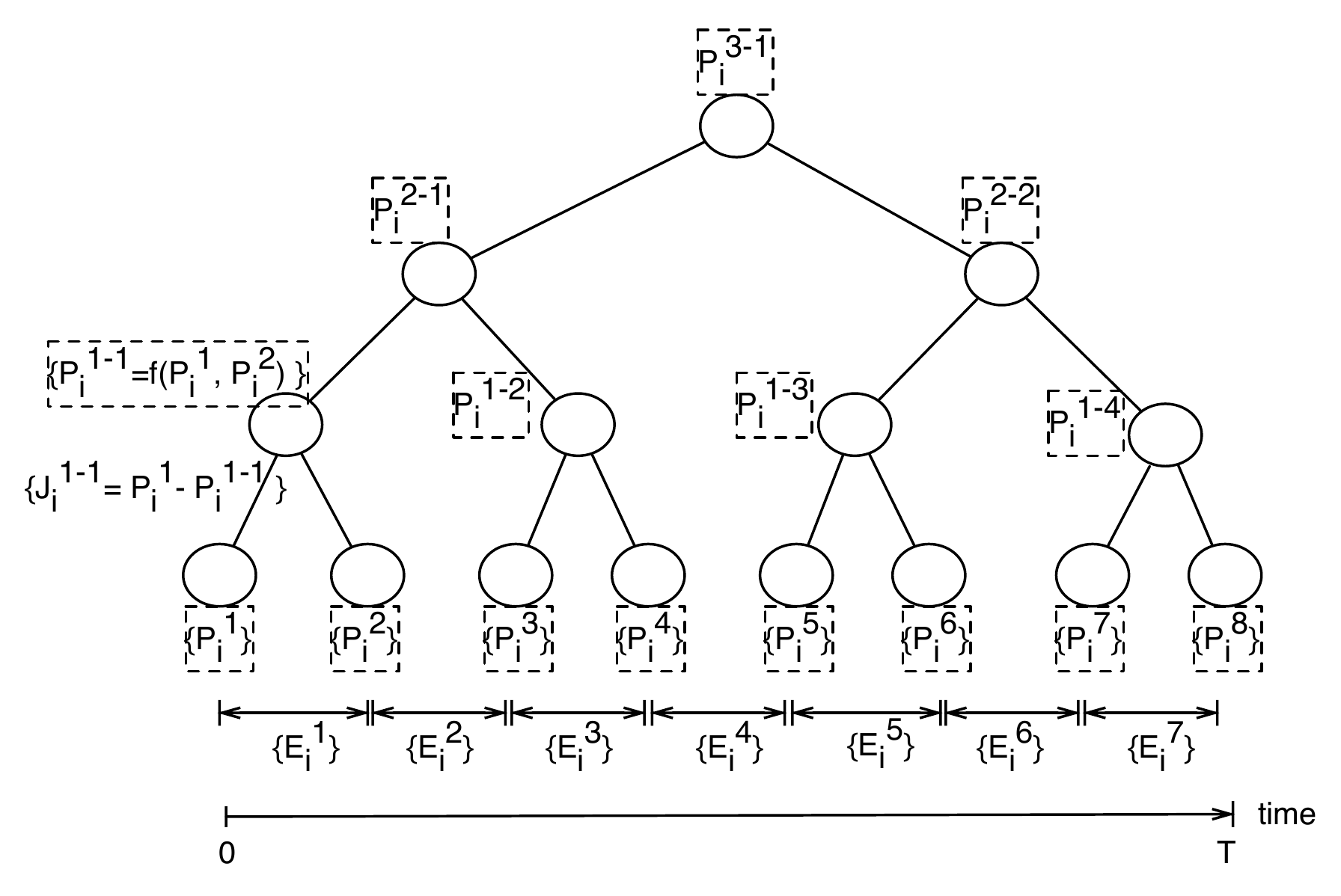}\label{fig:tgi-core}}\\
\subfloat[Version Chains]{\includegraphics[width = .17\textwidth]{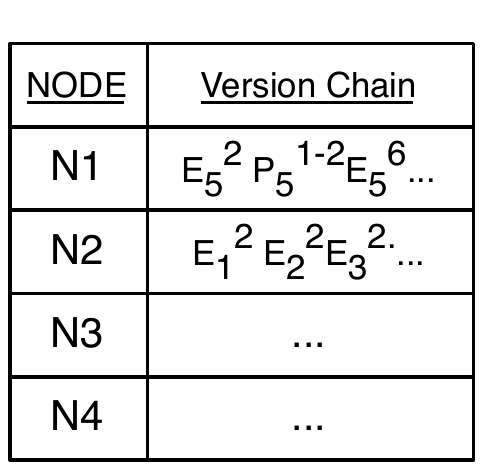}\label{fig:tgi-vc}}
\subfloat[Architecture]{\includegraphics[width = .32\textwidth]{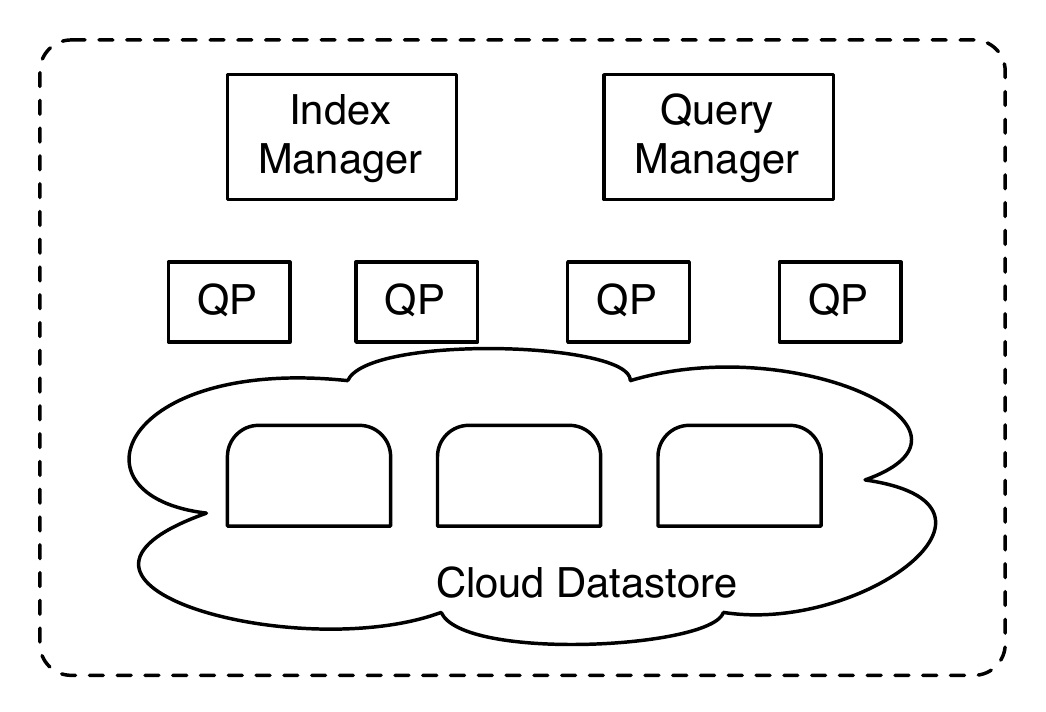}\label{fig:tgi-arch}}
\caption{Temporal Graph Index representation.}%: (a) core index with detlas; (b) node version tracking; (c) TGI component layout}
\label{fig:tgi-stuff}
\end{figure}

In short, the TGI stores \textit{deltas} or \textit{changes} in three different forms, as follows. The first one is the atomic changes in a chronological order through partitioned eventlists. This facilitates direct access to the changes that happened to a part or whole of the graph at specified points in time. Secondly, the state of nodes at different points in time is stored indirectly in form of the derived partitioned snapshot deltas. This facilitates direct access to the state of a neighborhood or the entire graph at a given time. Thirdly, a meta index stores node-wise pointers to the list of chronological changes for each node. This gives us a direct access to the changes occurring to individual nodes. Figure~\ref{fig:tgi-stuff}(a) shows the arrangement of eventlist, snapshot and derived snapshot partitioned deltas. Figure~\ref{fig:tgi-stuff}(b) shows a sample version chain.

TGI utilizes the concept of temporal consistency which was optimally utilized by DeltaGraph. However, it differs from DeltaGraph in two major ways. First, it uses a partitioning for eventlists, snapshots or deltas instead of a large monolithic chunks. Additionally, it maintains a list of version chain pointers for each node. The combination of these two novelties along with DeltaGraph's temporal compression generalizes the notion of entity-centric and time-centric indexing approaches in an efficient way. This can be seen by the qualitative comparison shown in Table~\ref{tab:deltacompare} as well as empirical results in Section~\ref{sec:experiments}.

%In this section, we claim that the combination of these direct access patterns provided by the TGI is the appropriate set for efficiently retrieving a wide variety of primitives on a historical graph.  

%\begin{figure}
%\begin{center}
%\includegraphics[width=0.9\linewidth, trim=0 0 0 10]{TGI.pdf}
%%%\vspace{-15pt}
%\caption{Temporal Graph Index representation: (a) TGI deltas - partitioned eventlists, snapshots and derived snapshots. The (dotted) bounded deltas are not stored; (b) Version Chains.}
%\label{fig:tgi-stuff}
%\end{center}
%%\vspace{-15pt}
%\end{figure}

%\begin{figure}[t]
%\subfloat[][TGI deltas - partitioned eventlists, \\snapshots and derived snapshots. The \\(dotted) bounded deltas are not stored.]{\includegraphics[width = .5\textwidth]{TGI.pdf}\label{fig:tgi-core}}
%%\subfloat[][Version Chains]{\includegraphics[width = .12\textwidth]{vc.pdf}\label{fig:tgi-vc}}
%%\subfloat[Architecture]{\includegraphics[width = .3\textwidth]{tgi-arch.pdf}\label{fig:tgi-arch}}
%\caption{Temporal Graph Index representation (XXXXX: Redraw).}%: (a) core index with detlas; (b) node version tracking; (c) TGI component layout}
%\label{fig:tgi-stuff}
%\end{figure}

%\begin{figure}
%\begin{center}
%\includegraphics[width=\linewidth]{TGI-core.pdf}
%\caption{}
%\label{fig:tgi-core}
%\end{center}
%\end{figure}
%
%\begin{figure}
%\begin{center}
%\includegraphics[width=\linewidth]{TGI-arch+VC.pdf}
%\caption{TGI: (a) Version chain lookup; (b) architecture.}
%\label{fig:tgi-arch-vc}
%\end{center}
%\end{figure}

\subsection{TGI: Design and Architecture}
In the previous subsection, we presented the logical description of TGI. We now describe the strategies for physical storage on a cloud which enables high scalability. 
In a distributed index, we desire that all graph retrieval calls achieve maximum parallelization through equitable distribution. A distribution strategy based on pure node-based key is good idea for snapshot style access, however, it is bad for a subgraph history style of access. A pure time-based key strategy on the other hand, has complementary qualities and drawbacks. 
An important related challenge for scalability is dealing with two different 
skews in a temporal graph dataset -- temporal and topological. They refer to the uneven density of graph activity over time and the uneven edge density across regions of the graph, respectively.
Another important aspect to note is that for a retrieval task, it is desirable that all the required micro-deltas on a particular machine be proximally located to minimize latency of lookups\footnote{In general, this depends on the underlying storage mechanism. While the physical placement of micro-deltas is irrelevant for a memory-based storage, it is significant for any disk-based storage due to seek times.}. 

\begin{figure}
\begin{center}
\includegraphics[width=\linewidth]{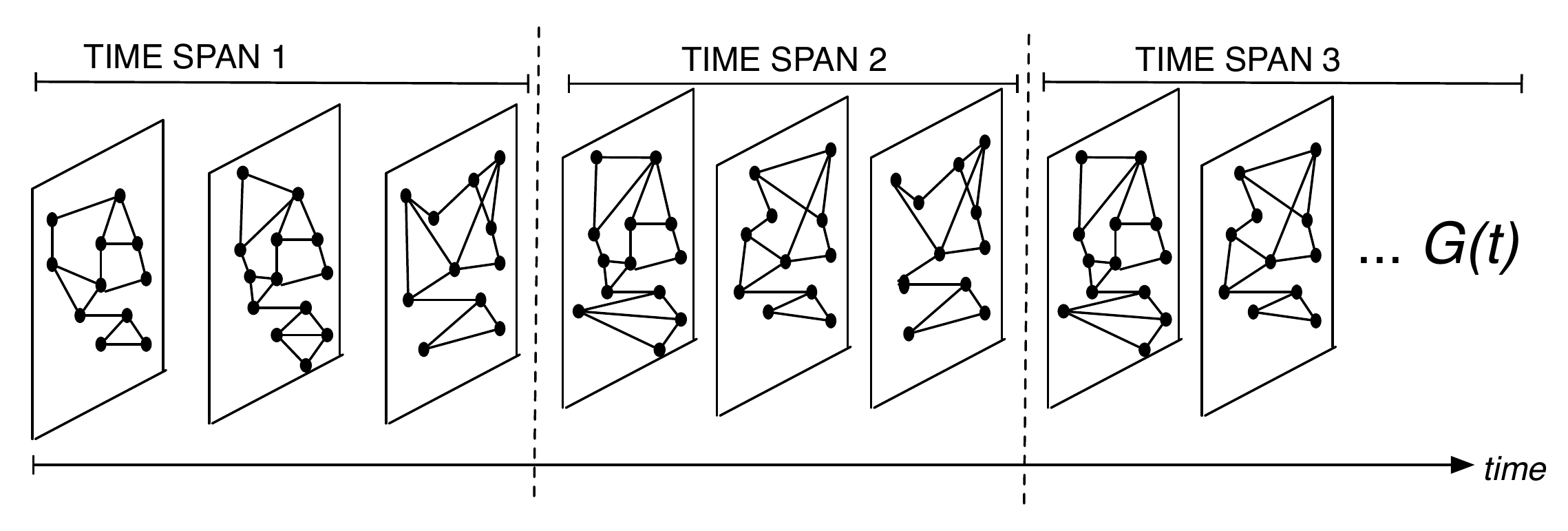}
\caption{The graph history is divided into non-overlapping periods of time. Such division is based on time intervals after which the locality-based graph partitioning is updated. It is also used as a partial key for data chunking and placement.}
\label{fig:timespans}
\end{center}
\end{figure}

Based on the above constraints and desired properties, we describe the physical layout of TGI as follows:
%So far, we have explained the core ideas behind TGI -- exploiting temporal consistency using a hierarchical index of finely partitioned snapshots, finely partitioned eventlists and a map for tracking the positions of nodes' changes, chronologically. We now discuss the key design and implementation aspects for scaling up TGI through cloud storage:
%\vspace{-5pt}
\begin{enumerate}
\itemsep0em
\item The entire history of the graph is divided into \textit{time spans}, keeping the number of changes to the graph consistent across different time spans, $f_t: e.time \rightarrow tsid$, where $e$ is the event and $tsid$ is the unique identifier for the time span. This is illustrated in Figure~\ref{fig:timespans}.
\item A graph at any point is horizontally partitioned into a fixed number of \textit{horizontal partitions} based upon a random function of the node-id, $f_h: nid \rightarrow sid$, where $nid$ is the node-id and $sid$ is unique identifier of for the horizontal partition.
\item The micro-deltas (including eventlists) are stored as a key-value pairs, where the delta-key is composed of \\$\{tsid, sid, did, pid\}$, where $did$ is a delta-id, and $pid$ is the partition-id of the micro-delta.
\item The placement-key is defined as a subset of the composite deltas key described above, as $\{tsid, sid\}$, which defines the chunks in which data is placed across a set of machines on a cluster. A combination of the $tsid$ and $sid$ ensure that a large fetch task, whether snapshot or version oriented, seeks data distributed across the cluster and not just one machine.
\item The micro-deltas are clustered by the delta key. The given order of the delta-key besides the placement-key elements, means that all the micro-partitions of a delta are stored contiguously, which makes it efficient to scan and read all micro-partitions belonging to a delta in a snapshot query. On the other hand, if the order of $did$ and $pid$ is reversed, it makes fetching a micro-partition across different deltas more efficient.
\end{enumerate}

%XXXXX: Summarize benefits of TGI design, either here or toward the end of the section.
Irrespective of a temporal or a topological skew in the graph, the index is
spread out across a cluster in a balanced manner. This also makes it possible
to fetch the graph primitives of large sizes in a naturally parallel manner.
For instance, a snapshot query would demand all micro-partitions for a specific
set of deltas in a particular timespan across all horizontal partitions. Given
an equitable distribution of the deltas across all machines of a cluster, we
retrieve the data in parallel on each storage machine, without a considerable
skew. 

\vspace{10pt}

\topic{Implementation:} TGI uses Cassandra for its delta storage. There are 5 tables that contain TGI data and metadata:\\
(1) \texttt{Deltas(tsid, sid, did, pid, dval)} table stores the deltas as described above, where \texttt{dval} contains 
serialized value of the micro-delta as a binary string. \\
(2) \texttt{Versions(nid, vchain)} consists of each node's version chain as a hash-table with keys 
for each timespan.\\ 
(3) \texttt{Timespans(tsid, start, end, checkpts, k, df)} stores, for each timespan, start and end times, a list of snapshot checkpoints, and arity. \\ %deleted differential function
(4) \texttt{Graph(start, end, events, tscount, gtype)} contains global information about the graph and TGI. \\
(5) \texttt{Micropartitions(nid, tsid, pid)} contains micro-delta partitioning information about nodes. It is not utilized in case of random partitioning.\\
The graph construction and fetching modules are written in Python, using Pickle and Twisted libraries for serialization and communication.

\vspace{10pt}

\topic{Architecture:} TGI architecture can be seen in Figure~\ref{fig:tgi-arch}, where \textit{Query Mananger (QM)} is responsible for planning, dividing and delegating the query to one or more \textit{Query Processors (QP)}. The QPs query the datastore in parallel and %perform the required set of instructions to 
process the raw deltas into the required result. Depending on the query specification, the distributed result is either aggregated at a particular QP (the QM) or returned to the client which made the request without aggregation. The \textit{Index Manager} is responsible for the construction and maintenance activities of the index. The cloud represents the distributed datastore.

\vspace{10pt}

\topic{Construction and Update:} The construction process involves three different stages.
First, we analyze the input data using the index construction parameters including
the timespan length (ts), number of horizontal partitions (ns), number of likely 
datastore nodes (m), eventlist size (l), and micro-delta partition size (psize). 
%Several secondary TGI parameters are derived using these.
%As
%a result, several secondary TGI parameters are derived and it is known that
%what are time span boundaries, timepoints for snapshot checkpoints, horizontal
%partitioning information, and other parameters required to construct the index.
In the second stage, the input data is split into horizontal partitions. In the
third stage, parallel construction workers of the IM work on separate
horizontal partitions, and build the index, a time span at a time. The process
of construction of each timespan is similar to that of DeltaGraph, albeit
more fine-grained due to delta partitioning and version chain construction as well. 
The TGI accepts updates of events in batches of timespan length. The update process involves
creating an independent TGI with the new events, and merging it with the original TGI.
The merger of TGIs involves updates of corresponding deltas, VC index and the metadata. 

%\topic{Construction:}  XXXXX:Expand this section with more details from the dissertation
%The construction process involves three different stages.
%First, we analyze the input data using the index construction parameters including
%the timespan length (ts), number of horizontal partitions (ns), number of likely 
%datastore nodes (m), eventlist size (l), and micro-delta partition size (psize). 
%In the second stage, the input data is split into horizontal partitions. In the
%third stage, parallel construction workers of the IM work on separate
%horizontal partitions, and build the index, a time span at a time. The process
%of construction of each timespan is similar to that of DeltaGraph, albeit
%more complex due to delta partitioning and version chain construction as well, and
%we refer the reader to our prior work for more details~\cite{icdepaper}.
%
\subsection{Dynamic Graph Partitioning}
\begin{figure}
\subfloat[Graph Snapshot.]{\includegraphics[width = .25\textwidth]{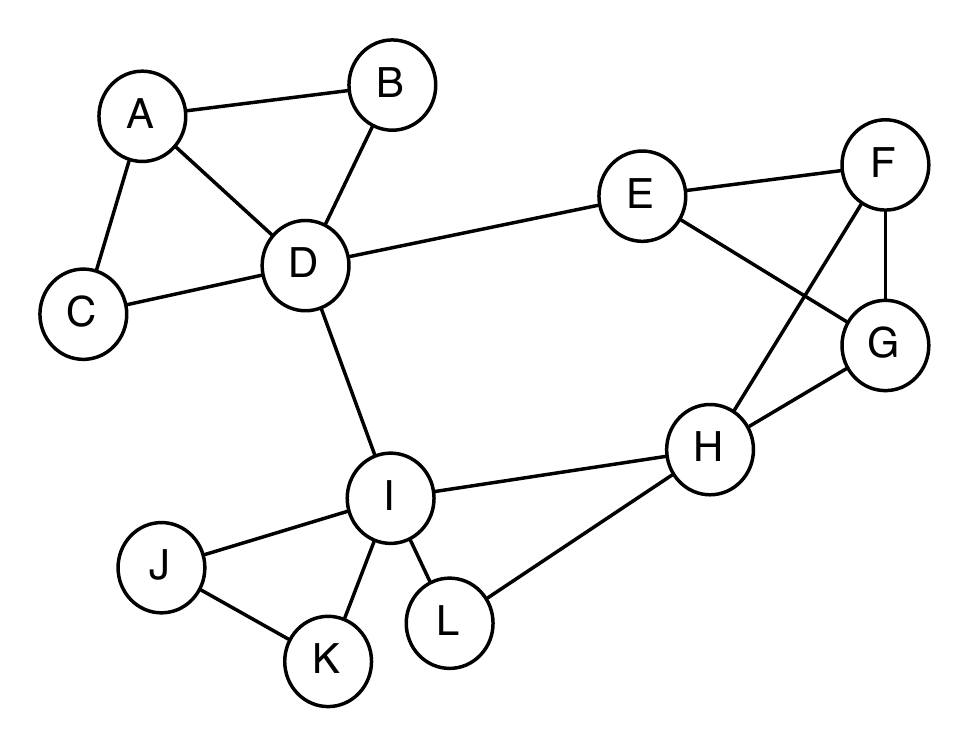}} 
\subfloat[Random partitioning of graph snapshot with high number of edge-cuts.]{\includegraphics[width = .25\textwidth]{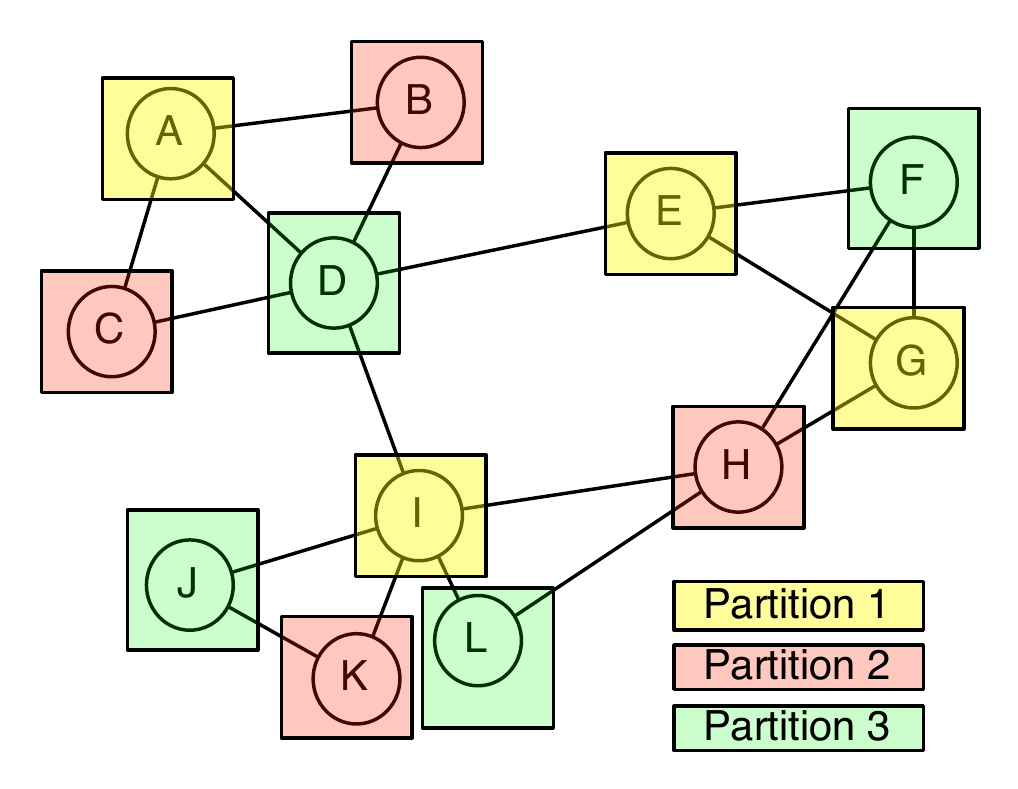}} \\
\subfloat[Min-cut partitioning of snapshot.]{\includegraphics[width = .25\textwidth]{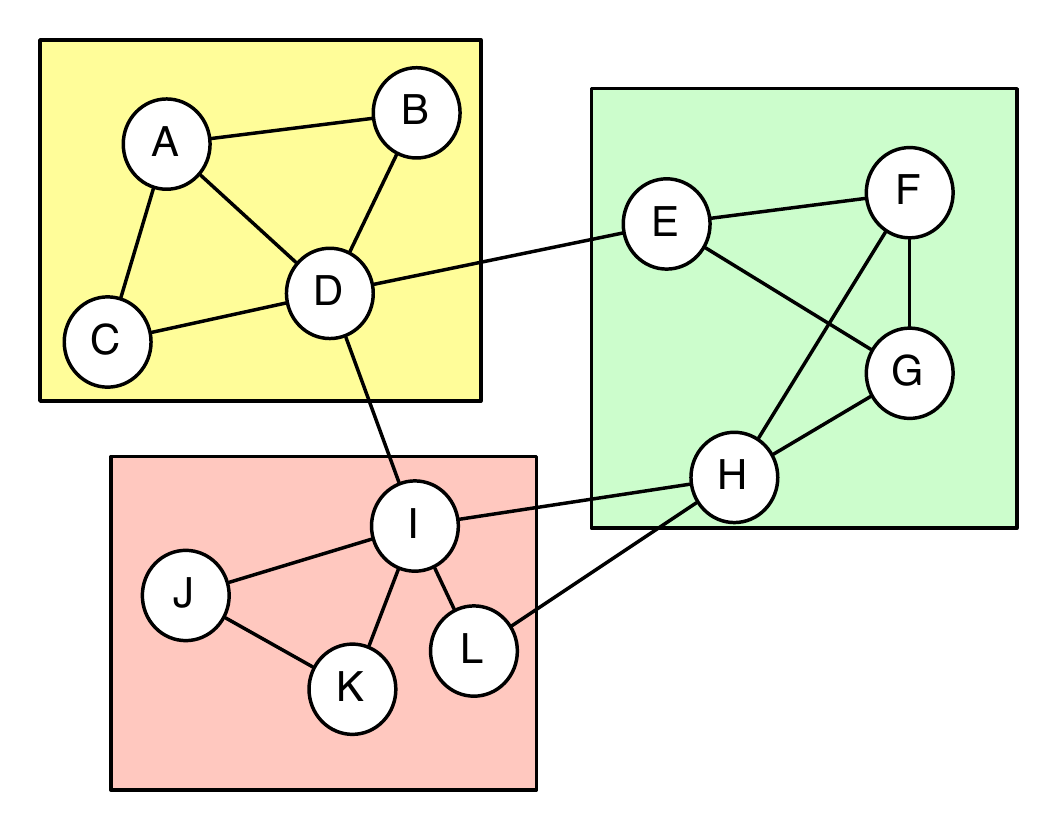}}
\subfloat[Min-cut partitioning of snapshot with edge-cut replication and auxiliary storage strategy.]{\includegraphics[width = .25\textwidth]{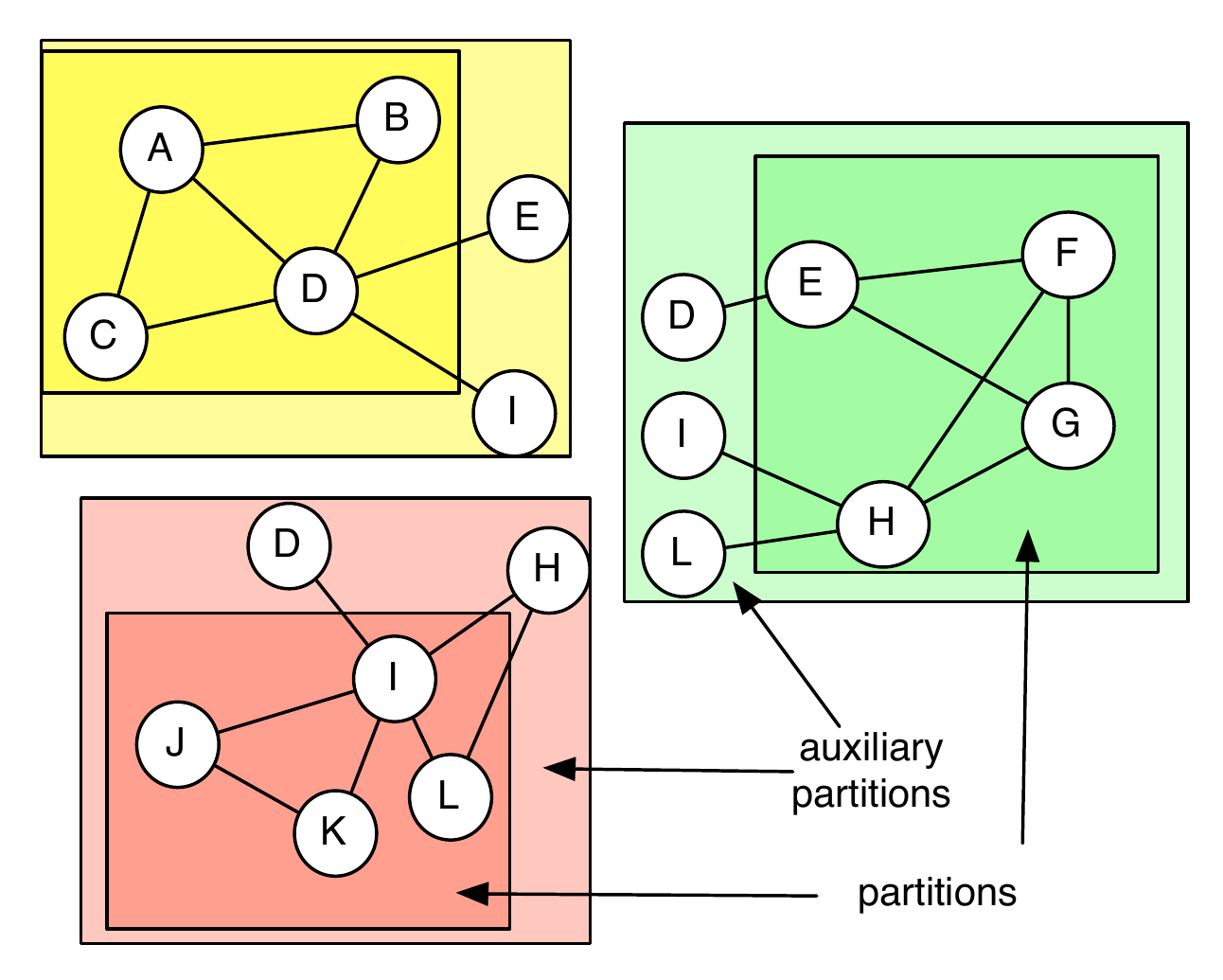}}
\caption{Graph partitioning using min-cut strategy along with 1-hop replication and the use of auxiliary micro-deltas improves 1-hop neighborhood retrieval performance without affecting the performance of snapshot or node retrieval.}
\label{fig:micrdeltasaux}
\end{figure}

Partitioning deltas into micro-deltas is an essential aspect of TGI and provides cheaper access to subgraph elements when compared to DeltaGraph or similar indexes. In a time-evolving graph, however, the size and topology of the graph change with time.
The key is to keep the size of each micro-delta (and each micro-eventlist) about the same and bounded 
by a number that dictates the latency for fetching a node or neighborhood. 
The two traditional approaches to partitioning a static graph are random (node-id hash-based) or locality-based (min-cut max-flow) partitioning. Random partitioning is simpler and involves minimal bookkeeping. 
However, since it loses locality, it is unsuitable for neighborhood-level granularity access. 
Locality-aware partitioning, on the other hand, preserves locality but incurs extra bookkeeping in form of a \{node-id:partition-id\} map. TGI is designed to work with either configuration as desired, as well as different partition size specifications. 
%Figure~\ref{fig:micrdeltasaux} pictorially depicts both approaches towards partitioning. Figure~\ref{fig:micrdeltasaux}(d) depicts replication to further speed-up 1-hop retrieval without jeopardizing the performance of other queries through a distinction between micro-deltas and auxiliaries.
TGI also supports replication of edge-cuts for further speed up of 1-hop neighborhoods. It uses a separate {\em auxiliary micro-delta} besides each micro-delta to store the replication, thereby preventing extra read cost for snapshot or node centric queries. This is illustrated in Figure~\ref{fig:micrdeltasaux}.

Locality-aware partitioning, however, faces an additional challenge with time-evolving graphs. % dynamic graphs and in the rest of this subsection, we provide insights on our solution.% as ageneral case of supporting efficient {\em k-hop neighborhood} retrieval.
With the change in size and topology of a graph, a partitioning deemed good (with respect to locality) at an instant may cease to be good at a later time.
A probable approach of frequent repartitioning over time would maintain partitioning quality, but leads to excessive amounts of bookkeeping, which in turn leads to degradation of performance while accessing different node or neighborhood versions.  
%Ideally, we could repartition or update a partitioning frequently so as to maintain a good partitioning
%\footnote{Optimal graph partitioning is NP-complete. We use the adjective ``good'' in reference to the quality of a partitioning, with respect to the respective graph snapshot partitioning heuristic used in a particular algorithm.}
%as per a particular notion of a graph partitioning heuristic. However, we also require a map for tracking partitions at each time we change the partitioning scheme.
%Maintaining and looking up that map as frequently as the changes in the graph is highly inefficient. 
Maintaining and looking up that map as frequently as the changes in the graph is highly inefficient. Hence, we divide the history of the graph into {\em time spans}, where
we keep the partitioning consistent within each time span, but perform it afresh it at the beginning of each new time span. 
This gives rise to two problems, described briefly as follows. 
Firstly, given a graph over time span, $T \in [t_s, t_e)$, find the graph partitioning that minimizes the edge cuts across all time points combined. Secondly, to determine the appropriate points for the end of a time span and the beginning of a new one, with respect to over all query performance. We discuss these problems below.

Static graph partitioning for an undirected and unweighted graph $G=(V, E)$ into $k$ partitions is defined as follows. 
Each node $v_i \in V$ is assigned a partition set $P_r$ such that $0 \le r < k$. The constraint is that $ \floor{|V| \over k} \le |P_r| \le \ceil{|V| \over k}$, i.e., the partitions are more or less equal in size. The number of edge cuts across partitions are intended to be minimized, i.e., 
%$C = \sum_{e \in E_c}{1} \forall e \in E_c s.t. $, where 
a count of all edges whose end points lie in different partitions. For a weighted graph, the edge cut cost is counted as the sum of the edge weights, which pushes stronger relationships (with higher edge weights) to be preferred for being in the same partition over
the lesser weighted ones. Also, in case of a node weighted graph, the partition set count can be determined using the node weight. Different graph partitioning algorithms work under these constraints using one or the other heuristic, as described before.

For a dynamic graph partitioning, we consider an edge and node weighted, undirected time evolving graph, without the loss of generality. Consider the following: graph $G^T=(V^\tau, E^\tau, W_E^\tau, W_N^\tau)$ where, $\tau \in [t_s, t_e)$, is the time range for which we 
find a single partitioning; $V^\tau, E^\tau, W_E^\tau, W_N^\tau$ are the set of vertices, edges, edge weights, node weights over time $\tau$, respectively.
Our partitioning strategy involves projecting the graph over time range $T$ to a single point in time using a \textit{time collapsing} function $\Omega$, there by reducing the graph $G^\tau$ to a static graph, $G_\tau = \Omega(G^\tau)$. The constraint on function $\Omega$ is that $G_\tau$ must contain all the vertices that existed in $G^\tau$ at least once in $G^\tau$. Using $G_\tau$, we can employ
static graph partitioning to find a suitable partitioning technique in the following manner.

The choice of $\Omega$ function determines how well the $G_T$ is a representation of $G^\tau$. Let us consider a few different options. (1) Median: consider the time point $t$ which is the median of the end points of $\tau$. The edges and weights in $G_\tau$ are the edge weights in $G^t$. (2) Union-Max: for an edge that existed at any time in $G^\tau$, we include it in $G_\tau$ such that its weight is the maximum value from all time points in $G^T$. (3) Union-Mean: for an edge that existed at any time in $G^\tau$, we include it in $G_\tau$, where its weight is the weighted average (time fraction) of the edge weights in $G^\tau$. Non existence of an edge during a time period counts as a $0$ contribution for the respective time period. (4). For any of the cases above, the node weight, $w_n$, can be defined independently of the edge set and edge weights. We consider three options as follows. (1) $w_n = 1$ for each nodes $n$ in $G_\tau$; (2) $w_n=degree(n)$ for each node $n$ in $G_T$; (3) $w_n=\mean{degree(n)}$, i.e., average degree over $\tau$.

Given these different heuristic combinations, we plan to study their empirical behavior and use the apparently most suitable one for TGI partitioning. The default TGI partitioning uses Union-Max for edge weights and uniform node weights.

We argue that this style of partitioning that involves first projecting a temporal graph to a static one, followed by conventional forms of static graph partitioning, is better than other conceivable alternatives. One such alternative way of doing it is to determine the partitioning at different time points in $\tau$, say $P^\tau$ and then reducing $P^\tau$ to $P_\tau$, a single partitioning scheme. This approach has the following major disadvantages. Firstly, the output partitions from a a static graph partitioning algorithm for two 
versions of graph $G^\tau$, say $G^1$ and $G^2$ are not aligned, even when the two snapshots are similar to a large extent. This is attributed to some degree of randomness associated with graph partitioning algorithms. This makes it infeasible to combine $P^1$ and $P^2$ in to a single result. Secondly, this approach is much more expensive compared to our approach, because it involves computing $\tau$ orders of partitions. Another alternative approach is to use one of the \textit{online graph partitioning} algorithms, which updates a partition set for a graph upon a small change in the graph. However, the output of such an approach only gives us partitioning schemes at different time points. The partitions across time are better aligned to each other than the previous approach, but we would still need to compute a combined partitioning from all available partitions, and the notion of time collapsing is inevitable. Secondly, the partitioning results from incremental graph partitioning are often inferior compared to the batch mode of partitioning for obvious reasons.

Determining the appropriate number and the exact boundaries of time-spans is another important issue. The need for creating higher number of time-spans and hence reducing the duration of a time-span is to maintain healthier partitioning. Let the hit taken on query latencies (assuming a certain query load Q) due to a subpar snapshot partitioning be, $f(T)$. This hit is generally incurred on k-hop queries, without replication, due to higher number of micro-delta seeks. In case of replication across partitions, the degree of replication increases with inferior partitioning, and leads to indirect impact on query latencies. On the other hand, there is need to create longer time-spans because the version queries require multiple micro-deltas, at different time points. Higher the changing number of partitions over query's time interval, say $t$, higher the query latency. Let us say that for an average query time interval (again, as per a specific query load), the gain due to longer time spans is, $g(T)$. The appropriate length of a average time-span hence is the solution of the maxima of $g(T) - f(T)$. In practice, uniform time-span length in numbers of the number of events is perhaps the most convenient. While the models of $f$ and $g$ are complex, a good number for size of $T$ can be observed empirically.

%\begin{figure}
%\subfloat[Graph Snapshot.]{\includegraphics[width = .24\textwidth]{mp1.pdf}} 
%\subfloat[Random partitioning of graph snapshot with high number of edge-cuts.]{\includegraphics[width = .24\textwidth]{mp2.pdf}} \\
%\subfloat[Min-cut partitioning of snapshot.][Min-cut partitioning of \\snapshot.]{\includegraphics[width = .24\textwidth]{mp3.pdf}}
%\subfloat[Min-cut partitioning of snapshot with edge-cut replication and auxiliary storage strategy.]{\includegraphics[width = .24\textwidth]{mp4.pdf}}
%\caption{Graph partitioning using min-cut strategy along with 1-hop replication and the use of auxiliary micro-deltas improves 1-hop neighborhood retrieval performance without affecting the performance of snapshot or node retrieval.}
%\label{fig:micrdeltasaux}
%\end{figure}

\subsection{Fetching Graph Primitives}
We briefly describe access methods for different graph primitives. The algorithms provided here use primitive TGI fetch methods whose description should be self-explanatory from their nomenclature.
%Different access methods perform the loading of the required graph primitives from the TGI. The query manager which receives a query, makes the appropriate query plan, partitions the query plan, most often by dividing the task by set of $sid$s. The fetch effort is federated to multiple query processors; however, it works under the assumption that all the deltas are present in a single cloud and hence a local delta fetch is not discriminated from a remote delta fetch. The fetch results can be requested in two different modes, distributed or aggregated. In the former, the results end up distributed in the memories of different query processors. In the latter one, they are aggregated on one of the designated query processors. We succinctly describe some of the access methods below. The details of the algorithms are omitted due to space constraints.

\topic{Snapshot Retrieval:} In snapshot retrieval, the state of a graph at a time point is retrieved. Given a time $t_s$, the query manager locates the appropriate time span $T$ such that $t_s \in T$, within which, it figures out the path from the root of the TGI to the leaf closest to the given time point. All the snapshot deltas, $\Delta_{s1}, \Delta_{s2}, \dots, \Delta_{sm}$, (i.e., all their micro-partitions) along that path and the eventlists from the leaf node to the time point, $\Delta_{e1}, \Delta_{e2}, \dots, \Delta_{en}$ are fetched and merged appropriately as: $\sum_{i=1}^m{\Delta_{si}} + \sum_{i=1}^n{\Delta_{ei}}$ (notice the order). This is performed across different query processors covering the entire set of horizontal partitions. The procedure for snapshot retrieval is specified in Algorithm~\ref{algo:sr}.

\begin{algorithm}[h!]
\caption{Snapshot Retrieval}\label{algo:sr}
\begin{algorithmic}[1]
\Procedure{GetSnapshot}{$t$}\Comment{Graph at time $t$}
   \State $t'\gets {\sf GetNearestPartTime}(t)$
      \State $K\gets {\sf GetNearestPartKeys}(t)$
      \State $D \gets {\sf GetDeltas}(K)$
      \State $g \gets \emptyset$
 \For{$d: D$} 
 \State $g \gets g + d$
  \EndFor
  \State $ B \gets {\sf GetEventLists}(t',t)$
   \For{$b: B$}
   \State $b \gets {\sf FilterByTime}(b, t',t)$ 
 \State $g \gets g +b$
  \EndFor
   \State \textbf{return} $g$\Comment{The snapshot}
\EndProcedure
\end{algorithmic}
\end{algorithm}

\topic{Node's history:} Retrieving a node's history during time interval, $[t_s, te)$ involves finding the state of the graph at point $t_s$, and all changes during the time range $(t_s, t_e)$. The first one is done in a similar manner to snapshot retrieval except the fact that we look up only a specific micro-partition in a specific horizontal partition, that the node belongs to. The second part happens through fetching the node's version chain to determine its points of changes during the given range. The respective eventlists are fetched and filtered for the given node. The procedure for node-history retrieval is specified in Algorithm~\ref{algo:nh}.

\begin{algorithm}
\caption{Node's History}\label{algo:nh}
\begin{algorithmic}[1]
\Procedure{GetNodeHistory}{$I$,$t_s$, $t_e$}

 \Comment{Node I's history for $t_s$ to $t_e$}
   \State $C\gets {\sf GetVC}(I)$
   \State $ C \gets {\sf FilterByTime}(C, t_s, t_e)$
    \State $D \gets {\sf GetDeltas}(C)$

      \State $I_N \gets \emptyset$
 %\For{$d: D$} 
% \If {$t(c) \ge t_s \wedge t(c) \le t_e$ }
 \State $D \gets {\sf FilterByTime}(D,t_s,t_e)$
 \State $D \gets {\sf FilterById}(D,I)$
  \For{d:D}
  \State $I_N \gets I_N \cup d$ 
 \EndFor
 %\EndIf
  %\EndFor
   \State \textbf{return} $I_N$\Comment{Node's history}
\EndProcedure
\end{algorithmic}
\end{algorithm}

\topic{k-hop neighborhood (static):} In order to retrieve the k-hop neighborhood of a node, we can proceed in two possible ways. One of them is to fetch the whole graph snapshot and filter the required subgraph. The other is to fetch the given node, and then determine its neighbors, fetch them, and recurse. It is easy to see that the performance of the second method will deteriorate fast with growing $k$. However for lower values, typically $k \le 2$, the latter is faster or at least as good, especially if we are using neighborhood replication as discussed in a previous subsection. In case of a neighborhood fetch, the query manager automatically fetches the auxiliary portions of deltas (if they exist), and if the required nodes are found, further lookup is terminated. Two different procedures for fetching a k-hop neighborhood are specified in Algorithm~\ref{algo:khop1} and Algorithm~\ref{algo:khop2}, respectively.

\begin{algorithm}
\caption{Node's k-Hop Neighborhood (1)}\label{algo:khop1}
\begin{algorithmic}[1]
\Procedure{GetNodeKHopNeigh1}{$I$,$t_s$}

 \Comment{Node I's k-hop neighborhood at $t$}
 \State $g \gets {\sf GetSnapshot}(t)$
 \State $ C \gets \{I\}$
 \State $ R \gets \{I\}$
 
 \For{p:1 to k}
 \State{$S \gets \emptyset$}
 \For{ r:R}
 \State {$N \gets {\sf GetNeighbors}(g, r)$}
 \State $C \gets C \cup N$
 \State $S \gets S \cup N$
 \EndFor
\State{ $R \gets S$}
\EndFor
 \State $g' \gets {\sf FilterByID}(g, C)$
   \State \textbf{return} $g'$\Comment{Node's k-hop}
\EndProcedure
\end{algorithmic}
\end{algorithm}

\begin{algorithm}
\caption{Node's k-Hop Neighborhood (2)}\label{algo:khop2}
\begin{algorithmic}[1]
\Procedure{GetNodeKHopNeigh2}{$I$,$t$}

 \Comment{Node I's k-hop neighborhood at $t$}
\State $N \gets  {\sf GetNode}(I,t)$
\State $M \gets { \sf GetNeighbors}(N)$
\State $G \gets \emptyset$
\For{r: 1 to k}
\State $L \gets \emptyset$
\For{m:M}
\If{$m \in G$}
\State $N \gets {\sf GetNode(m)}$
\State $G \gets G + R$
\State $L \gets L \cup  {\sf GetNeighbors}(m)$
\EndIf
\EndFor
\State{$M \gets L$}
\EndFor

   \State \textbf{return} $G$\Comment{k-hop neighborhood}
\EndProcedure
\end{algorithmic}
\end{algorithm}

\topic{Neighborhood evolution:} Neighborhood evolution queries can be posed in two different ways. First, requesting all changes for a described neighborhood, in which case the query manager fetches the initial state of the neighborhood followed by the events indicating the change. Second, requesting the state of the neighborhood at multiple specific time points. This translates to the retrieval of multiple single neighborhoods fetch tasks. Algorithm~\ref{algo:1hophist} specifies the procedure to fetch one hop neighborhood history. The general k-hop evolution process can be seen as a combination of the 1-hop evolution procedure along with the k-hop (static) neighborhood retrieval.

\begin{algorithm}[h]
\caption{Node's 1-Hop History}\label{algo:1hophist}
\begin{algorithmic}[1]
\Procedure{GetNode1HopHistory}{$I$,$t_s$, $t_e$}

 \Comment{Node I's 1-hop history for $t_s$ to $t_e$}
\State{ $H \gets {\sf GetNodeHistory}(I, t_s, t_e)$}
\State{$G \gets \{H\}$}
\State{$S \gets \emptyset$}
\Comment{S is a set of pairs $<$Node,time-range$>$}
\For{h:H}
\State{$S \gets {\sf UpdateNeighborInfo}(S,h)$}
\EndFor
\For{s:S}
\State{$G \gets G \cup s$}
\EndFor
   \State \textbf{return} $G$
   \Comment{Node's 1-hop history}
\EndProcedure
\end{algorithmic}
\end{algorithm}

%XXX: Summarize the design and architecture contribution of TGI: (a) Various types of queries + (b) Scalability

%\topic{Architecture:}
%The TGI is a \textit{distributed} index where the deltas are spread across a set of shards as seen in Figure~\ref{fig:tgi-arch-vc}(b). All incoming fetch requests are federated to individual shards and need based communication happens between shards through a central \textit{fetch coordinator}, asynchronously. The end of the fetch process on all individual shards means the end of the overall fetch process for a certain fetch task. Based upon a configuration with the fetch coordinator (which can be overridden in the fetch request) the fetched output is aggregated on one machine or kept distributed. The \textit{index manager} creates the deltas and updates them upon updates in the graph. The \textit{layout manager} determines the physical placement of the deltas at all times.

\section{Analytics Framework}
\label{sec:taf}
%In order to perform a wide range of analytics on temporal graphs, there is a
%need for two of the following. Firstly, an medium to conveniently express a
%rich set of analytical needs, perhaps in a declarative fashion. Secondly, a
%scalable platform to execute those analytical tasks. 
In this section, we describe the \textit{Temporal Graph Analysis Framework (TAF)},
%consisting of a \textit{temporal graph analysis library}, 
that enables programmers to express and execute
complex analytical tasks on time-evolving graphs. %Using a set of different
%operators, users can specify complex temporal graph analysis objectives. 
We present details of the novel model of computation, including a library of temporal graph operators and operands (exposed through Python and Java APIs); we also present the details of implementation 
on top of Apache Spark, which enables scalable, parallel, in-memory execution. Finally, we describe TAF's coordination with TGI to provide a complete ecosystem for historical graph management and analysis.% of historical graph data. 
%To the best of our knowledge, none of the existing graph data management systems provide such capabilities around temporal graph analysis.

\subsection{Temporal Graph Analysis Library}
\label{subsec:tafcore}
\begin{figure}
\begin{center}
\includegraphics[width=0.9\linewidth, trim= 0 5 0 10]{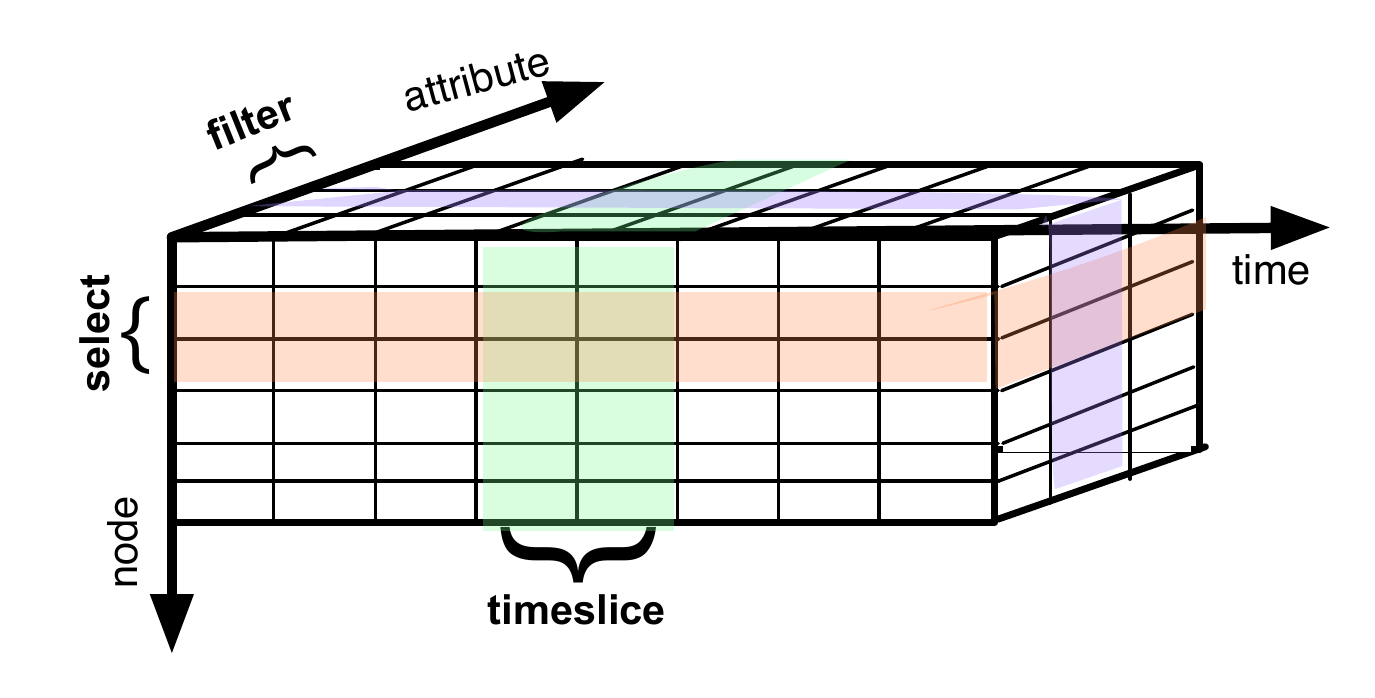}
%%\vspace{-15pt}
\caption{SoN: A set of nodes can be abstracted as a 3 dimensional array with temporal, node and attribute dimensions.}
\label{fig:son}
\end{center}
%\vspace{-15pt}
\end{figure}

In this section, we describe a set of operators for analyzing large historical graphs. At the heart of this library is a data model where we view the historical graph as a \textit{set of nodes or subgraphs evolving over time}. % The edges and attributes are properties of the nodes. 
The choice of temporal nodes as a primitive helps us describe a wide range of fetch and compute operations 
in an intuitive manner. More importantly, it provides us an abstraction to parallelize computation. 
The \textit{temporal nodes} and \textit{set of temporal nodes} bear a correspondence to \textit{tuples} and \textit{tables} of the relational algebra, as the basic unit of data and the prime operand, respectively.
%\vspace{-2pt}
\topic{Operands:} The two central data types are defined below:% that most operators operate upon.
%\vspace{-2pt}
\begin{definition}[Temporal Node] A \textit{temporal node} (NodeT), $ N^T$, is defined as a sequence of all and only the states of a node $ N$ over a time range, $T=[t_s, t_e)$. All the $k$ states of the node must have a valid time duration $T_i$, such that $\cup_i^k T_i = T$ and $\cap_i^k T_i = \phi$.
\end{definition}
%\vspace{-10pt}
\begin{definition}[Set of Temporal Nodes] A \textit{SoN}, is defined as a set of $r$ temporal nodes $\{ N_1^T, N_2^T \dots N_r^T \}$ over a time range, $T=[t_s, t_e)$, as depicted in Figure~\ref{fig:son}.
\end{definition}

The {\em NodeT} class provides a range of methods to access the state of the node at various time points, 
including: {\tt getVersions()} which returns the different versions of the node as a list of static nodes (NodeS),
{\tt getVersionAt()} which finds a specific version of the node given a timepoint, {\tt getNeighborIDsAt()} which
returns IDs of the neighbors at the specified time point, and so on.

A {\em Temporal Subgraph (SubgraphT)} generalizes NodeT and captures a sequence of the states of a subgraph (i.e., a set of nodes and edges among them) over a period of time. 
Typically the subgraphs correspond to $k$-hop neighborhoods around a set of nodes in the graph. An analogous {\tt getVersionAt()} function can be
used to retrieve the state of the subgraph as of a specific time point as an in-memory {\tt Graph} object (the user program must ensure that any graph
object so created can fit in the memory of a single machine). A Set of Temporal Subgraphs (SoTS) is defined analogously to SoN as a set of temporal subgraphs.

\topic{Operators:} Below we discuss the important temporal graph algebra operators supported by our system.\\
%%\vspace{-10pt}
\begin{enumerate}
%\vspace{-5pt}
%\itemsep-1em
\item {\bf \em Selection} accepts an SoN or an SoTS along with a boolean function on the nodes or the subgraphs, and returns an SoN or SoTS. 
Selection performs \textit{entity-centric filtering} on the operand, and does not alter temporal or attribute dimensions of the data. \\
\item {\bf \em Timeslicing} accepts an SoN or an SoTS along with a timepoint (or time interval) $t$, finds the state of each of individual nodes or subgraphs in the operand as of $t$, and returns it as another SoN or SoTS, respectively (SoN/SoTS can represent sets of static nodes or subgraphs as a well). The operator can accept a list of timepoints as input and return a list.\\
%selects one or more timeslices of a given operand, i.e., a given a time $t_1$, for each 
%
%i.e., $\tau_{t_1}(G) = \{g: t_1 \in g.time, g \in \pi_{ALL-A_{t},A_{t}=t_1}(G) \}$, i.e., all the nodes that were valid at timepoint $t_1$ are included in the output and the valid time for all output nodes is $t_1$. Multiple parameters, say $t_1$ and $t_2$ produce a collection of two different nodesets or two different snapshots. No parameter means produce every valid snapshot.
\item {\bf \em Graph} accepts an SoN and returns an in-memory Graph object containing the nodes in the SoN (with only the edges whose both endpoints are in the SoN). %The edge sets of the nodes in SoN are pruned to reflect only the relationships that exist in the context of the given SoN. 
An optional parameter, $t_p$, may be specified to get a GraphS valid at time $t_p$.\\% Note that the return Graph object is not distributed, and hence the user must make sure that it can fit in the memory of a single machine.
%\item Append ($\pi^{-1}$): This appends a list of attributes to an SoN. It is most usable when the programmer wants to preserve the result of a computation as an attribute of the nodes.
%\item AddNode ($\alpha$): This operator adds a new temporal node to an SoN.
\item {\bf \em NodeCompute} is analogous to a {\em map} operation; it takes as input an SoN (or an SoTS) and a function, and applies the function to all the individual nodes (subgraphs) and returns the results as a set.\\
\item {\bf \em NodeComputeTemporal}. Unlike {\tt NodeCompute}, this operator takes as input a function that operates on a static node (or subgraph) in addition to an SoN (or an SoTS); for each node (subgraph), it returns a sequence of outputs, one for each different state (version) of that node (or subgraph). Optionally, the user may specify another function (NodeComputeDelta) that operates on the delta between two versions of a node (subgraph), which the system can use to compute the output more efficiently. An optional parameter is a method describing points of time at which computation needs to be performed; in the absence of it, the method will be evaluated at all the points of change.\\
\item {\bf \em NodeComputeDelta} operator takes as input: (a) a function that operates on a static node (or subgraph) and produces an output quantity, (b) an SoN (or an SoTS) like \\{\tt NodeComputeTemporal}, (c) a function that operates on the following: a static node (or subgraph), some auxiliary information pertaining to that state of the node (or subgraph), the value of the quantity at that state, and an \textit{update} (event) to it. This operator returns a sequence of outputs, one for each state of the node (or subgraph), similar to \\{\tt NodeComputeTemporal}. However, the method of computation in this method is different because it updates the computed quantity for each version incrementally instead of computing it afresh. An optional parameter is the method describing points of time at which to base the comparison. An optional parameter is a method describing points of time at which computation needs to be performed; in the absence of it, the method will be evaluated at all the points of change.\\
%This operator is analogous to a {\em map} operation; it takes as input an SoN (or an SoTS) and a function, and applies the function to all the individual nodes (subgraphs) and returns the results as a set.
%%or subgraphsis meant for a user defined function to be defined for a temporal node. It takes as input a NodeT and returns a pair of node-id and another return object as specified by the UDF. NodeComputeStatic($\mu_S$) is a variation of this graph that operates upon a static node.
%\item NodeComputeDelta($\mu_{\delta}$): This operator implements user defined logic for a temporal node as a substitute to MapNode. However, instead of the compute logic on temporal nodes, the user provides logic to compute on a static node as well as a delta in form of events.   
%\item GraphCompute($\mu_{G}$): This operator helps specify the logic of computation on a temporal graph through user defined functions on temporal nodes. GraphStaticCompute is a similar specific case where the operand is a static graph.
\item {\bf \em Compare} operator takes as input two SoNs (or two SoTSs) and a scalar function (returning a single value), computes the function value over all the individual components, and returns the differences between the two as a set of {\em (node-id, difference)} pairs. This operator tries to abstract the common operation of comparing two different snapshots of a graph at different time points. A simple variation of this operator takes a single SoN (or SoTS) and two timepoints as input, and does the compare on the timeslices of the SoN as of those two timepoints. An optional parameter is the method describing points of time at which to base the comparison. \\
\item {\bf \em Evolution} operator samples a specified quantity (provided as a function) over time to return evolution of the quantity over a period of time. An optional parameter is the method describing points of time at which to base the evolution.\\
%\item Aggregation($\gamma$): It is an abstraction of specific statistical aggregation functions such as \textit{max, min, saturates, sum, average, count,} etc.
%\item NodeReduce($\rho$): This is a general reduce operator that can perform arbitrary types of user defined aggregation functions per node.
\item {\bf \em TempAggregation} abstractly represents a collection of temporal aggregation operators such as {\tt Peak}, {\tt Saturate}, {\tt Max}, {\tt Min}, and {\tt Mean} over a scalar timeseries. The aggregation operations are used over the results of temporal evaluation of a given quantity over an SoN or SoTs. For instance, finding ``times at which there was a {\it peak} in the network density'' is used to find eventful timepoints of high interconnectivity such as conversations in a cellular network, or high transactional activity in a financial network.
\end{enumerate}

%Our system supports several other manipulation and mutation operations that we omit here for lack of space.

\subsection{System Implementation}
The library is implemented in Python and Java and is built on top of the Spark API. %Table~\ref{tab:taf-api}, shows some of the key function calls. 
The choice of Spark provides us with an efficient in-memory cluster compute execution platform, circumventing dealing with the issues of data partitioning, 
communication, synchronization, and fault tolerance. %Also, we were able to quickly implement several components by building on top of methods provided in the expressive Spark API. 
%In this subsection, we will talk about the several important aspects of the system implementation for the analytics framework.
We provide a GraphX integration for utilizing the capabilities of the Spark based graph processing system for static graphs.

The key abstraction in Spark is that of an RDD, which represents a collection of objects of the same type, stored across a cluster.
SoN and SoTS are implemented as RDDs of NodeT and SubgraphT respectively (i.e., as {\tt RDD<NoteT>} and {\tt RDD<SubgraphT>}). 
The in-memory graph objects may be implemented using any popular graph representation, specially the ones that support useful libraries on top.
%we use the open-source TinkerGraph implementation~\cite{tinkergraph}, but
%other in-memory graph implementations could be used as well. TinkerGraph supports the Blueprints Graph API, and thus the graph objects that are created can be
%manipulated or analyzed using the Gremlin query language and other software packages built on top of Blueprints. 
We now describe in brief the implementation details for NodeT and SubgraphT, followed by details of the incremental computational operator, and the parallel data fetch operation. 

\begin{figure}[t]
\centering
%\vspace{-10pt}
\subfloat[Finding node with highest local clustering coefficient]{\includegraphics[width =0.5\textwidth, trim = 0 10 0 0]{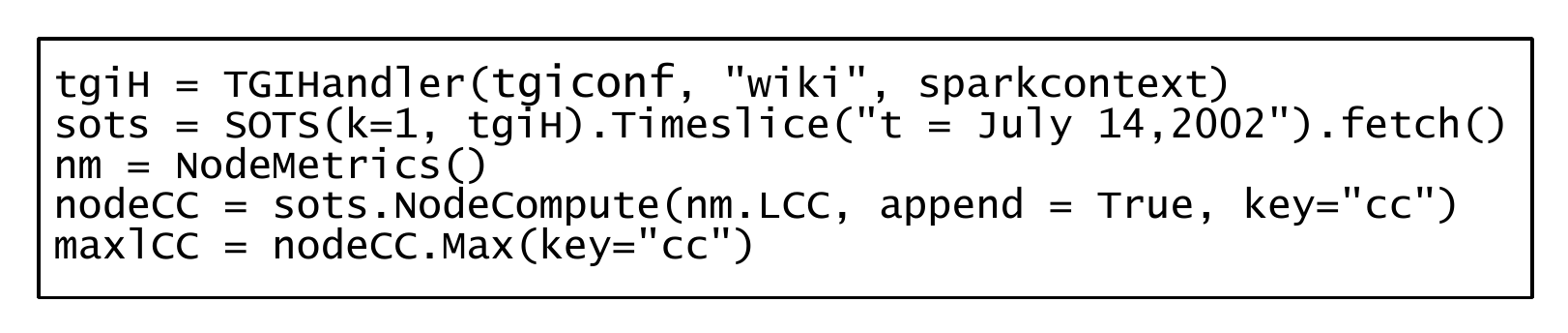}} \\
%\vspace{-10pt}
\subfloat[Comparing two communities in a network]{\includegraphics[width = 0.5\textwidth, trim = 0 10 0 0]{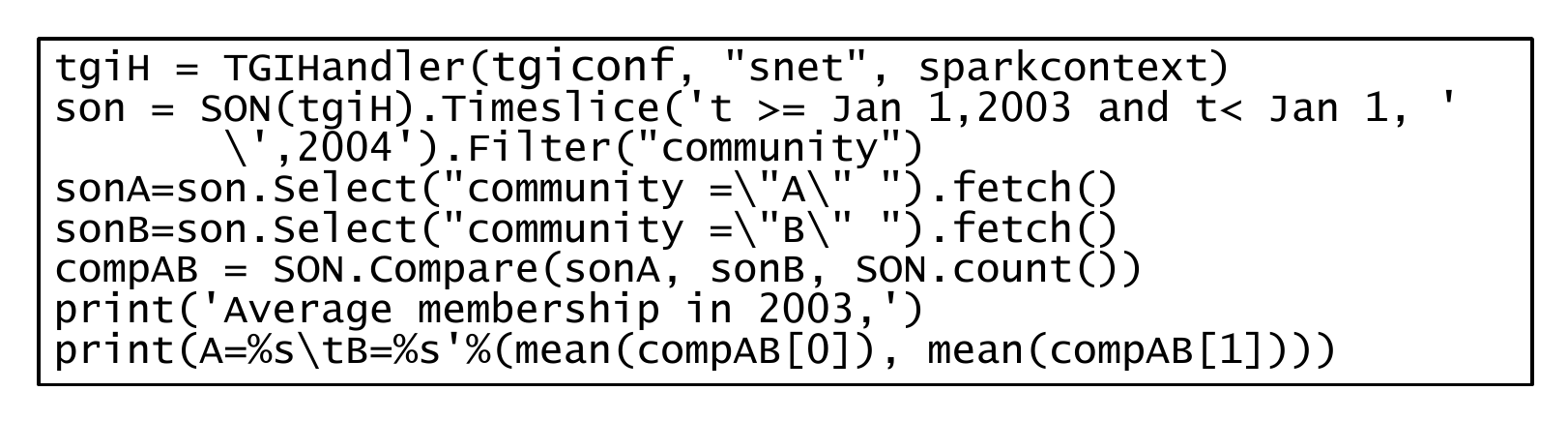}}\\
%\vspace{-10pt}
\subfloat[Evolution of network density]{\includegraphics[width = 0.5\textwidth, trim = 0 10 0 0 ]{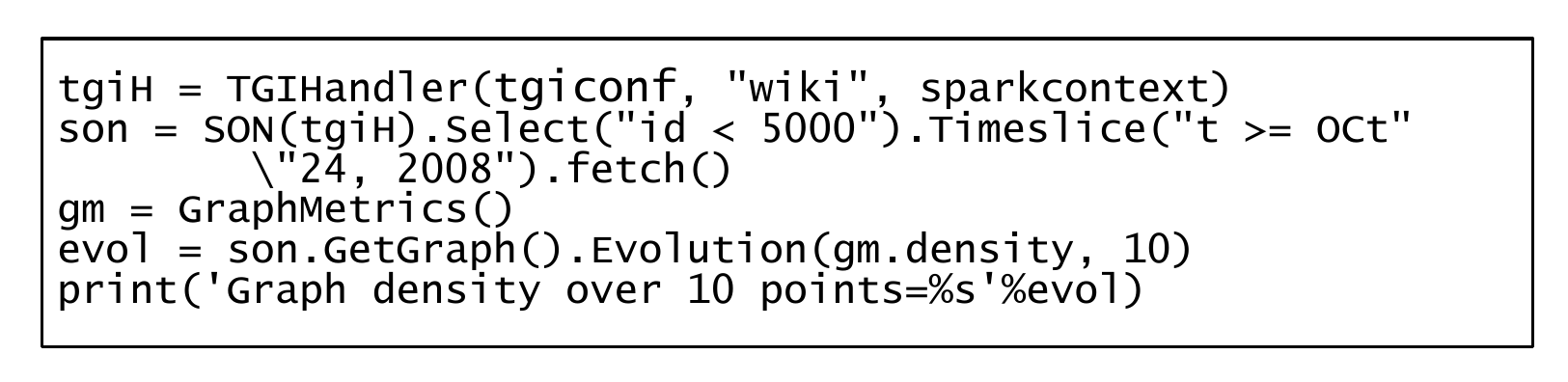}}
%%\vspace{-5pt}
\caption{Examples of analytics using the TAF Python API.}
\label{fig:taf-code-ex}
%\vspace{-15pt}
\end{figure}

Figure~\ref{fig:taf-code-ex} shows sample code snippets for three different analytical tasks -- (a) finding the node with the \textit{highest clustering coefficient} in a historical snapshot; (b) \textit{comparing  different communities} in a network; (c) finding the \textit{evolution of network density} over a sample of ten points.

%\vspace{10pt}

\topic{NodeT and SubgraphT:}
A set of temporal nodes is represented with an \texttt{RDD} of \texttt{NodeT} (temporal node). A temporal node contains the information for a node during a specified time interval. The question of the appropriate physical storage of the \texttt{NodeT} (or \texttt{SubgraphT}) structure is quite similar to storing a temporal graph on disk such as the one using a DeltaGraph or a TGI, however, in-memory instead of disk. Since NodeT is fetched at query time, it is preferable to avoid creating a complicated index, since the cost to create the index at query time is likely to offset any access latency benefits due to the index. An intuitive guess based upon examination of certain temporal analysis tasks is that its access pattern is most likely going to be in a chronological order, i.e., the query requesting the subsequent versions or changes, in order of time. Hence, we store \texttt{NodeT} (and \texttt{SubgraphT}) as an initial snapshot of the node (or subgraph), followed by a list of chronologically sorted events. It provides methods such as \texttt{GetStartTime()}, \texttt{GetEndTime()}, \texttt{GetStateAt()}, \texttt{GetIterator()}, \texttt{Iter-\\ator.GetNextVersion()}, \texttt{Iterator.GetNextEvent()}, and so on. We omit the details of these methods as their functionality is apparent from the nomenclature.

%\vspace{10pt}

\begin{figure}[h!]
\centering
\subfloat[Using {\tt NodeComputeTemporal}]{\includegraphics[width =0.5\textwidth, trim= 0 10 0 15]{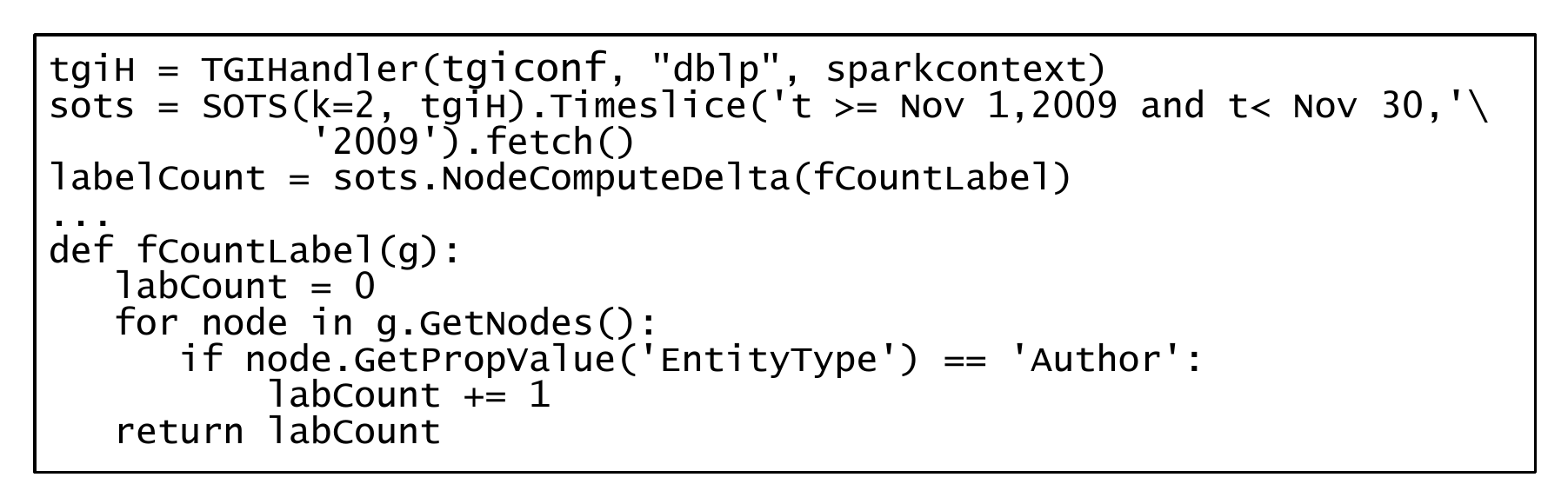}}\\

\subfloat[Using {\tt NodeComputeDelta}]{\includegraphics[width = 0.50\textwidth, trim= 0 10 0 20]{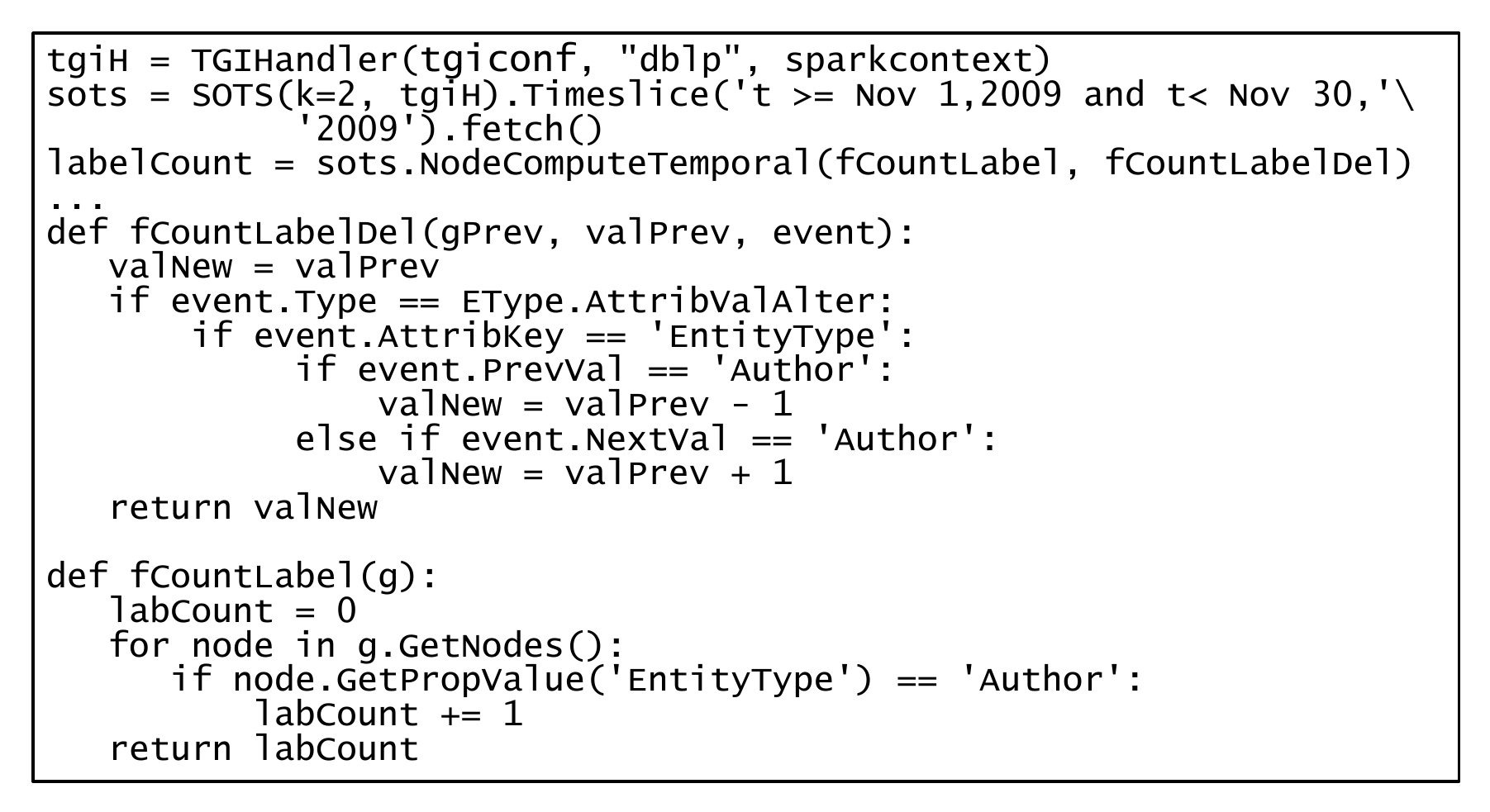}}
\caption{Incremental computation using different methods compute counts of nodes with a specific label in subgraphs over time.}
\label{fig:computedelta_taf}
\end{figure}

\topic{NodeComputeDelta:}
\texttt{NodeComputeDelta} evaluates a quantity over each NodeT (or SubgraphT) using two supplied methods, $f()$ which computes the quantity on a state of the node or subgraph, and, $f_\Delta()$, which updates the quantity on a state of the node or subgraph for a given set of event updates. Consider a simple example of finding the fraction of nodes with a specific attribute value in a given \texttt{SubgraphT}. If this were to be performed using \\\texttt{NodeComputeTemporal}, the quantity will be computed afresh on each new version of the subgraph, which would cost $\mathcal{O}(N.T)$ operations where $N$ is the size of the operand (number of nodes) and $T$ is the number of versions. However, using the incremental computation, each new version can be processed in constant time after the first snapshot, which adds up to, $\mathcal{O}(N + T)$. While performing the incremental computation, the corresponding $f_\Delta ()$ method is expected to be defined so as to evaluate the nature of the event -- whether it brings about any changing the output quantity or not, i.e., a scalar change value based upon the actual event and the concerned portions of the state of the graph, and also update the auxiliary structure, if used. Code in Figure~\ref{fig:computedelta_taf} illustrates the usage of {\tt NodeComputeTemporal} and {\tt NodeComputeDelta} in a similar example.

Consider a somewhat more intricate example, where one needs to find counts of a small pattern {\em over time} on an {\tt SoTS}, such as finding the occurrence of a subgraph pattern in the data graph's history. In order to perform such pattern matching over long sequences of subgraph versions, it is essential to maintain certain inverted indexes which can be looked up to answer in constant time whether an event has caused a change in the answer from a previous state or caused a change in the index itself, or both. Such inverted indexes, quite common to subgraph pattern matching, are required to be updated with every event; otherwise, with every new event update, we would need to look up the new state of the subgraph afresh which would simply reduce it to performing non-indexed subgraph pattern matching over new snapshots of a subgraph at each time point, which is a fairly expensive task. In order to utilize a constantly updated set of indices, the auxiliary information, which is a parameter and a return type for $f_\Delta()$, can be utilized.
Note that such an incremental computational operator opens up possibilities for using a considerable amount of algorithmic work available in literature on online and streaming graph query evaluation, respectively, to be applied to historical graph analysis. For instance, there is work on pattern matching in streaming~\cite{wang2009continuous, gao2014continuous} and incremental computing~\cite{fan2013incremental, varro2006incremental} contexts, respectively.

%Note that such an incremental computational operator opens up possibilities of utilizing a large body of algorithmic work in online and streaming graph query evaluation for the purpose of graph analytics. %For instance, there is work on pattern matching in streaming~\cite{wang2009continuous, gao2014continuous} and incremental computing~\cite{fan2013incremental, varro2006incremental} contexts, respectively.

%\begin{figure}
%\centering
%%%\vspace{-1pt}
%\subfloat[Patten to be  matched]{\includegraphics[width =0.11\textwidth]{IncrPattern.pdf}}
%%%\vspace{-1pt}
%\subfloat[Ongoing indexes need to be maintained to verify the pattern]{\includegraphics[width = 0.37\textwidth]{IncrPatternIndex.pdf}}\\
%%%\vspace{-1pt}
%
%\caption{Pattern matching in temporal subgraphs.}
%\label{fig:patt_taf}
%%\vspace{-10pt}
%\end{figure}

\begin{figure}[h]
%\vspace{-1pt}
\centering
\subfloat[Specifying the start, end and middle point of {\tt SON} for an {\tt Evolution} query.]{\includegraphics[width =0.5\textwidth, trim=0 10 0 16]{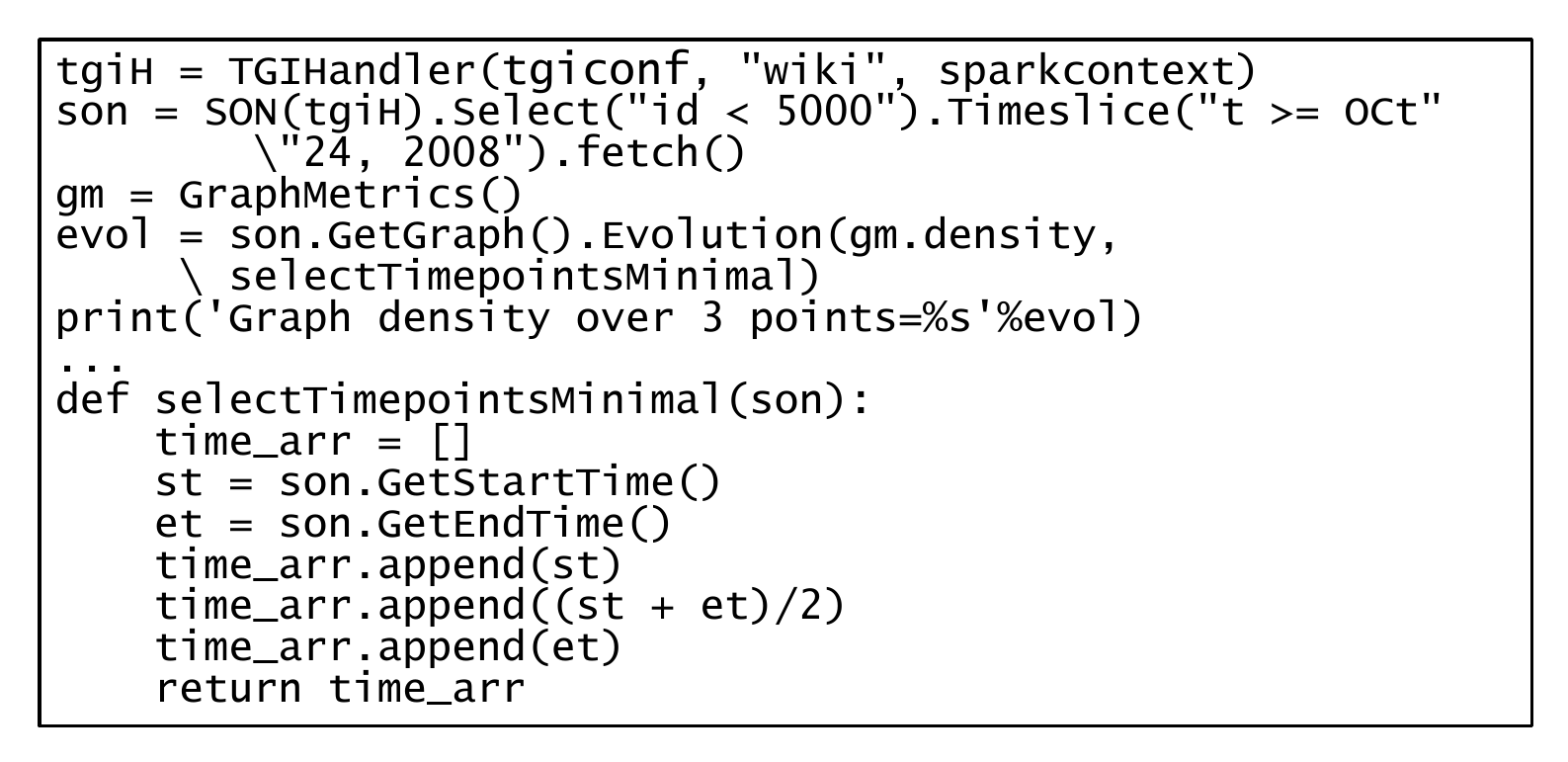}}\\
\subfloat[Specifying all change points in two {\tt SON}'s for a {\tt Compare} query.]{\includegraphics[width = 0.5\textwidth, trim=0 10 0 10]{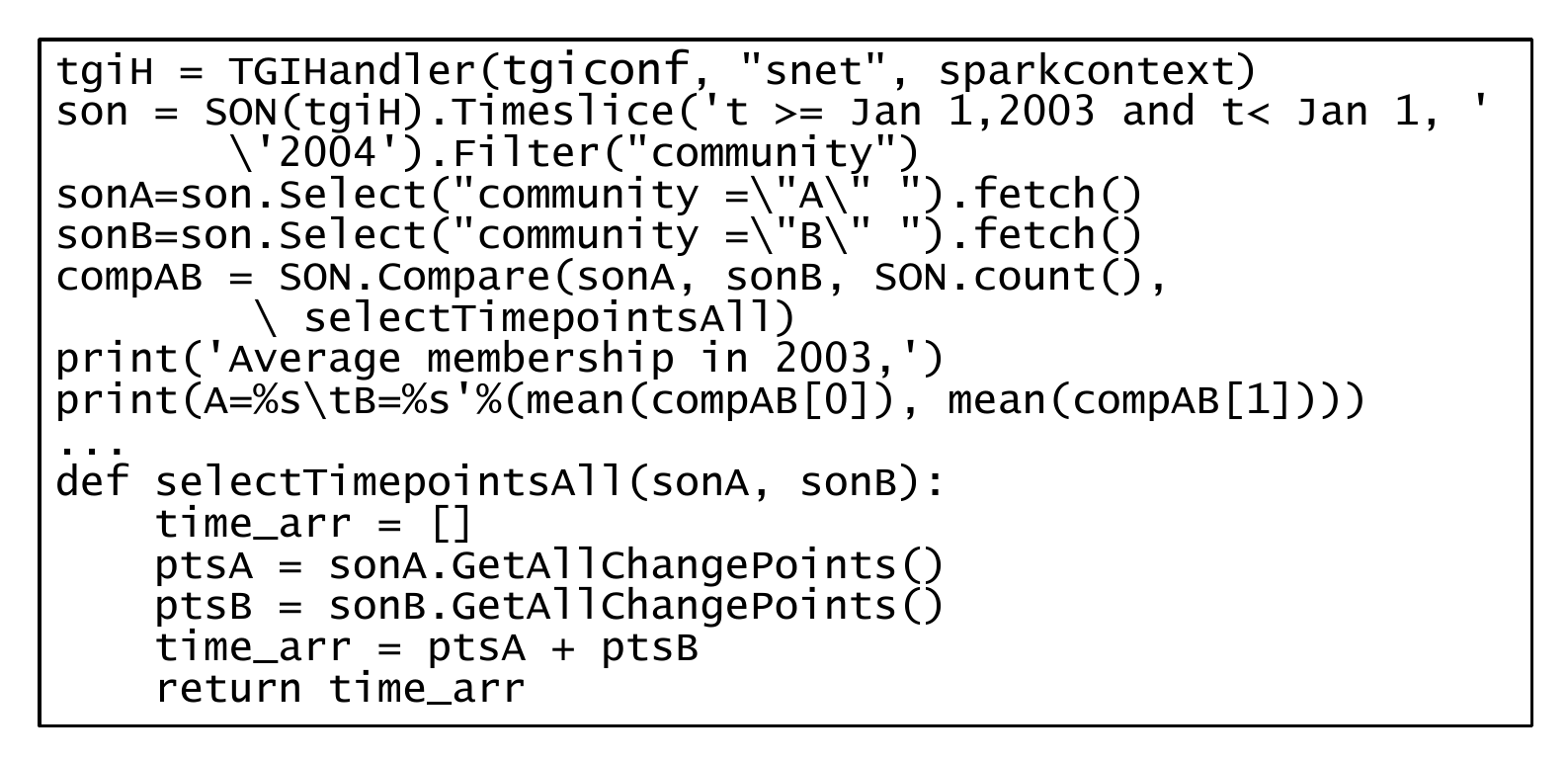}}
\caption{Using the optional timepoint specification function with evolution and comparison queries.}
\label{fig:timepts_taf}
%%\vspace{-5pt}
\end{figure}

\topic{Specifying interesting time points:}
In the \text{map}-oriented \text{version} operators on an {\tt SoN} or an {\tt SoTS}, the time points of evaluation, by default, are all the points of change in the given operand. However, a user may choose to provide a definition of which points to select. This can be as simple as returning a constant set of timepoints, or based on a more complex function of the operand(s). Except the {\tt Compare} operator, which accepts two operands, other operators allow an optional function, which works on a singe temporal operand; the compare accepts a similar function that operates on two such operands. Two such examples can be seen in Figure~\ref{fig:timepts_taf}.

%\vspace{40pt}

\begin{figure}[h]
\vspace{0pt}
\begin{center}
\includegraphics[width=\linewidth, trim = 10 0 10 0]{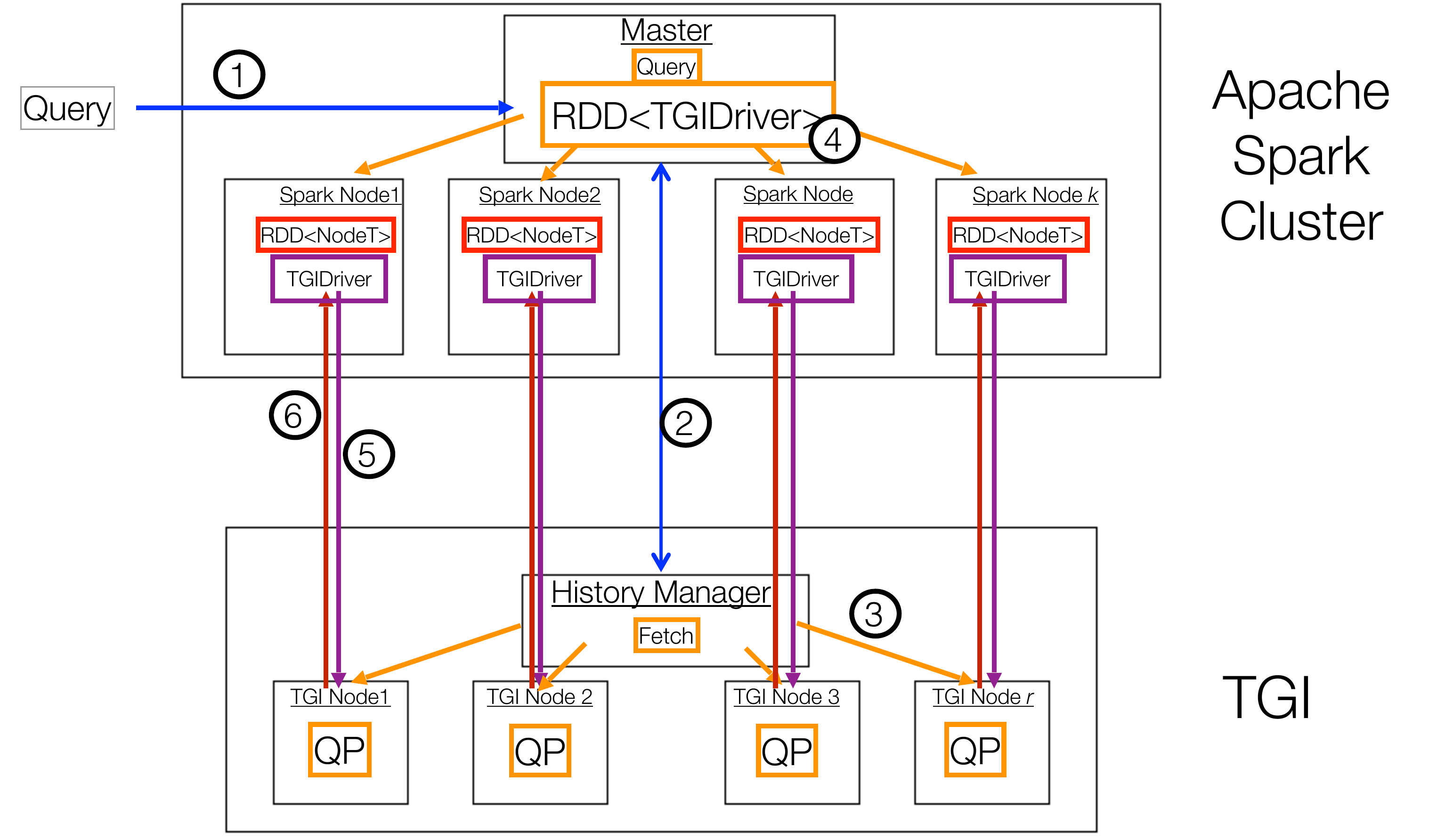}
\caption{A pictorial representation of the parallel fetch operation between the TGI cluster and the analytics framework cluster. The numbers in circles indicate the relative order of events and the arrowheads indicate the direction of flow.}
\label{fig:parallel-fetch}
%\vspace{-5pt}
\end{center}
\end{figure}

\topic{Data Fetch:}
In a temporal graph analysis task, we first need to instantiate a \textit{TGI connection handler} instance. It contains configurations such as address and port of the TGI \textit{query manager host}, \textit{graph-id}, and a \textit{SparkContext} object. Then, a SON (or SOTS) object is instantiated by passing the reference to the TGI handler, and any query specific parameters (such as k-value for fetching 1-hop neighborhoods with SOTS). The next few instructions specify the semantics of the graph to be fetched from the TGI. This is done through the commands explained in Section~\ref{subsec:tafcore}, such as the \texttt{Select}, \texttt{Filter}, \texttt{Timeslice}, etc. However, the actual retrieval from the index doesn't happen until the first statement following the specification instructions. A \texttt{fetch()} command can be used explicitly to tell the system to perform the fetch operation. Upon the \texttt{fetch()} call, the analytics framework sends the combined instructions to the query planner of the TGI, which translates those instructions into an optimal retrieval plan. This prevents the system from retrieving large amounts of data from the index that is a superset of the required information and prune it later. 

%Figure~\ref{fig:fetch-opt} shows a couple of such translations.
%\begin{figure}
%\begin{center}
%\includegraphics[width=\linewidth]{FetchOpt.pdf}
%\caption{Translation of fetch instructions specified in the TAF to fetch instructions in the TGI.}
%\label{fig:fetch-opt}
%\end{center}
%\end{figure}

The analytics engine runs in parallel on a set of machines, so does the graph index. The parallelism at both places speeds up and scales both the tasks. However, if the retrieval graph at the TGI cluster was aggregated at the Query Manager and sent serially to the master of the analytical framework engine after which it was distributed to the different machines on the cluster, it would create a space and time bottleneck at the Query Manager and the master, respectively, for large graphs. In order to bypass this situation, we have designed a parallel fetch operation, in which there is a direct communication between the nodes of the analytics framework cluster and the nodes of the TGI cluster. This happens through a protocol that can be seen in Figure~\ref{fig:parallel-fetch}. The protocol is briefly described in the following ordered steps:
%\\1. Analytics query containing fetch instructions is received by the TAF master. \\
%2. A handshake between the TAF master and TGI query manager is established. The latter receives the fetch instructions and the former is made aware of the active TGI query processor nodes.\\
%3. Parallel fetch starts at the TGI cluster.\\
%4. The TAF master instantiates a TGIDriver instance at each its cluster machines wrapped in a RDD.\\
%5. Each node at the TAF performs a handshake with one or more of the TGI nodes.\\
%6. Upon the end of fetch at TGI, the individual TGI nodes transfer a portion of the SoN to an RDD on the corresponding TAF nodes.

\begin{enumerate}
%%\vspace{-5pt}
%\itemsep0em 
\item Analytics query containing fetch instructions is received by the TAF master. 
\item A handshake between the TAF master and TGI query manager is established. The latter receives fetch instructions and the former is made aware of the active TGI query processor nodes.
\item Parallel fetch starts at the TGI cluster.
\item The TAF master instantiates a TGIDriver instance at each of its cluster machines wrapped in a RDD.
\item Each node at the TAF performs a handshake with one or more of the TGI nodes.
\item Upon completion of fetch at TGI, the individual TGI nodes transfer the SoN to an RDDs on the corresponding TAF nodes.
\end{enumerate}

%\begin{table}[htdp]
%\caption{A list of some important methods with the TAF library.}
%\begin{center}
%\begin{tabular}{|l|p{5cm}|}
%\hline
%Method & Description \\
%\hline
%\hline
%Select & Select the nodes of interest from index or another SoN\\
%\hline
%Filter & Filter or attach the node attributes of interest referring to another SoN\\
%\hline
%Timeslice & Select a specified part of SoN for a period or time or a snapshot\\
%\hline
%Fetch & Tells the engine to actually fetch the graph from the index\\
%\hline
%MapNode & Uses a UDF to operate on each temporal node in SoN\\
%\hline
%MapNodeDelta & Uses two UDFs, one to specify functionality on a node's version, another on node delta\\
%\hline
%Append & Appends an SoN with another or a Tuple with matching dimensions\\
%\hline
%NodeReduce & Aggregates results of a map by node-id.\\
%\hline
%GetRDD & Gives control of the core RDD of the SoN to the programmer\\
%\hline
%\end{tabular}
%\end{center}
%\label{tab:taf-api}
%\end{table}%

%\topic{Discussion:}
%TBD. Talk about TAF data model. How it presents us with an abstraction that is easy to parallelize+ express both node-centric vs graph centric stuff. 
%How it gives a platform to perform optimizations such as Map v MapDelta. 
%Also, we have a question of distributed vs single machine stuff graph exec().

\section{Experimental Evaluation}
\label{sec:experiments}

%The efficacy of HGS is judged by two measures. First, ``does its functionality enable analysts to specify a wide range of meaningful objectives, i.e., analytical tasks for historical graphs?''. Second, ``whether the solutions provided by the system are efficient in terms of time and other resources, and significantly better than the naive approached''. 
In this section, we empirically evaluate the efficiency of TGI and TAF.
To recap, TGI is a persistent store for entire histories of large graphs, that
enables fast retrieval for a diverse set of graph primitives -- snapshots, subgraphs, and nodes at past time points or across intervals of time.
We primarily highlight the performance of TGI across the entire spectrum of retrieval primitives. We are not aware of a baseline that may compete with TGI across all or a substantial subset of these retrieval primitives.
Specialized alternatives such as DeltaGraph for snapshot retrieval is highly unsuitable for node or neighbor version retrieval; a version centric index may be specialized for node-version retrieval but is highly unsuitable for snapshot or neighborhood-version style retrieval. Also note that TGI generalizes all the known approaches including those two; using appropriate parameter configurations, it can even converge to any specific alternative.
Secondly, we demonstrate the scalability of TGI design through experiments on parallel fetching for large and varying data sizes.
%versions of nodes or neighborhoods, etc. 
%The range of different retrieval primitives pose two distinct
%trade-offs in indexing and retrieval. First, the size of a subgraph can vary
%from a node to an entire graph. Second, the history of a subgraph to be retrieved can range
%from one time point to the graph's entire time span. We show how TGI captures these trade-offs. 
%Additionally, we report results for demonstrating its scalability over large data and heavily parallel fetching. 
Finally, we also report experiments demonstrating computational scalability of the TAF for a graph analysis task, as well as the benefits of our incremental computational operator.

\topic{Datasets and Notation:} We use four datasets: (1) Wikipedia citation
network consisting of 266,769,613 edge addition events from Jan 2001 to Sept
2010. At its largest point, the graph consists of 21,443,529 nodes and
122,075,026 edges; (2) We augment Dataset 1 by adding around 333 million
synthetic events which randomly add new edges or delete existing edges over a period of
time, making a total of 700 million events; (3) Similarly, we add
733 million events, making the total around 1 billion events; (4) Using a Friendster gaming network snapshot, 
we add synthetic dates at uniform intervals to 500 million events with a total 
of approximately 37.5 million nodes and 500 million edges.

Following key parameters that are varied in the experiments: data store machine count ($m$), 
replication across dataset ($r$), 
number of parallel fetching clients ($c$), 
eventlist size ($l$), snapshot or eventlist partition size ($ps$), and 
Spark cluster size ($m_a$).

We conducted all experiments on an Amazon EC2 cluster. Cassandra ran on machines containing 4 cores and 15GB of available memory. We did not use row caching and the actual memory consumption was much lower that the available limit on those machines. Each fetch client ran on a single core with up to 7.5GB available memory. The machines with TAF nodes running Spark workers ran on a single core and 7.5GB of available memory each.

\begin{figure}
\begin{center}
\includegraphics[width=\linewidth]{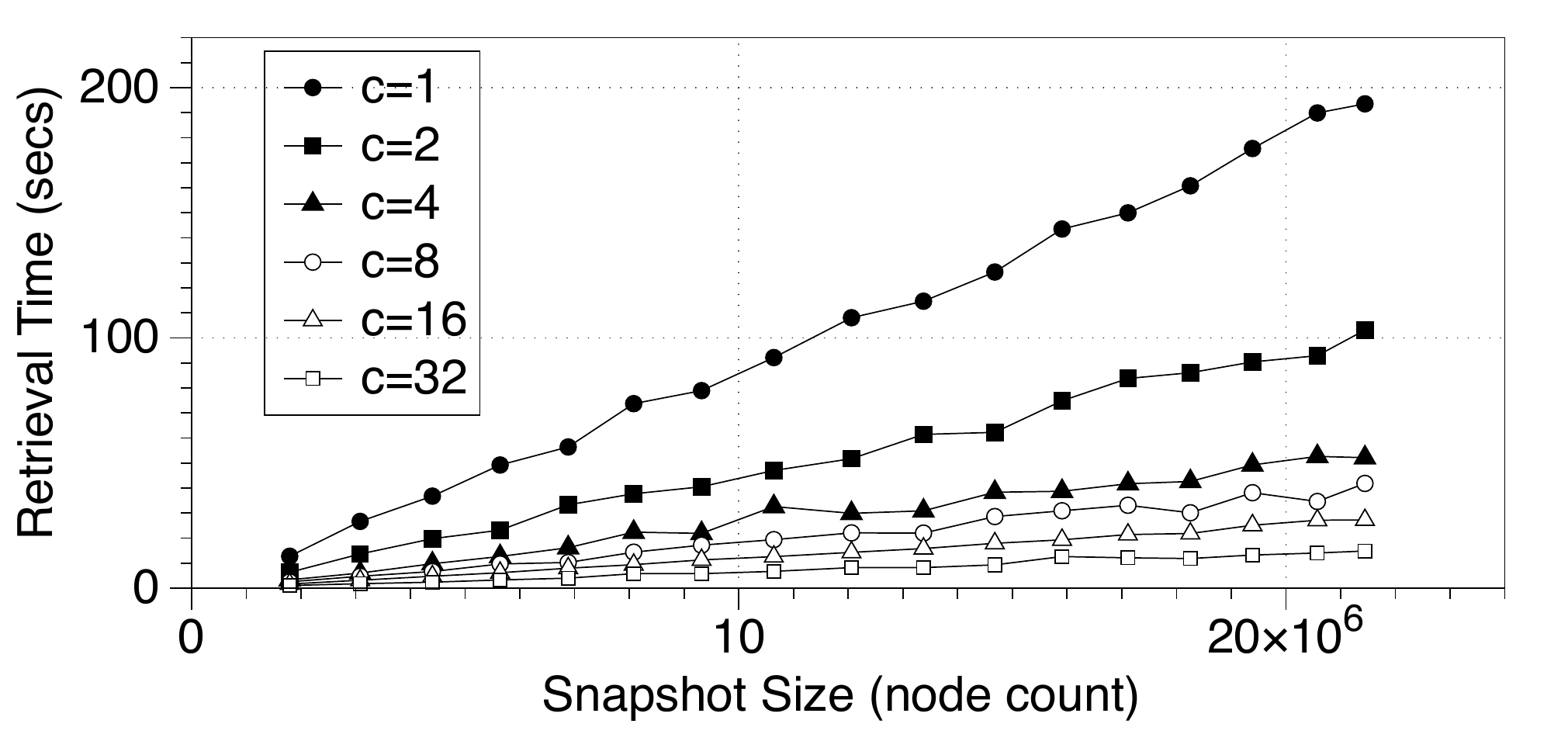}
\caption{Snapshot retrieval times for varying parallel fetch factor ($c$), on Dataset 1; $m=4$; $r=1, ps=500$.}
%\vspace{-12pt}
\label{fig:ss1}
\end{center}
%\vspace{-15pt}
\end{figure}

\begin{figure*}[h!]
%%\vspace{-10pt}
%\subfloat[Fetching and postprocessing times.]{\includegraphics[width = .33\textwidth]{ss-breakup.pdf}\label{fig:breakup}} 
\subfloat[Compressed vs. uncompressed delta storage.]{\includegraphics[width = .33\textwidth]{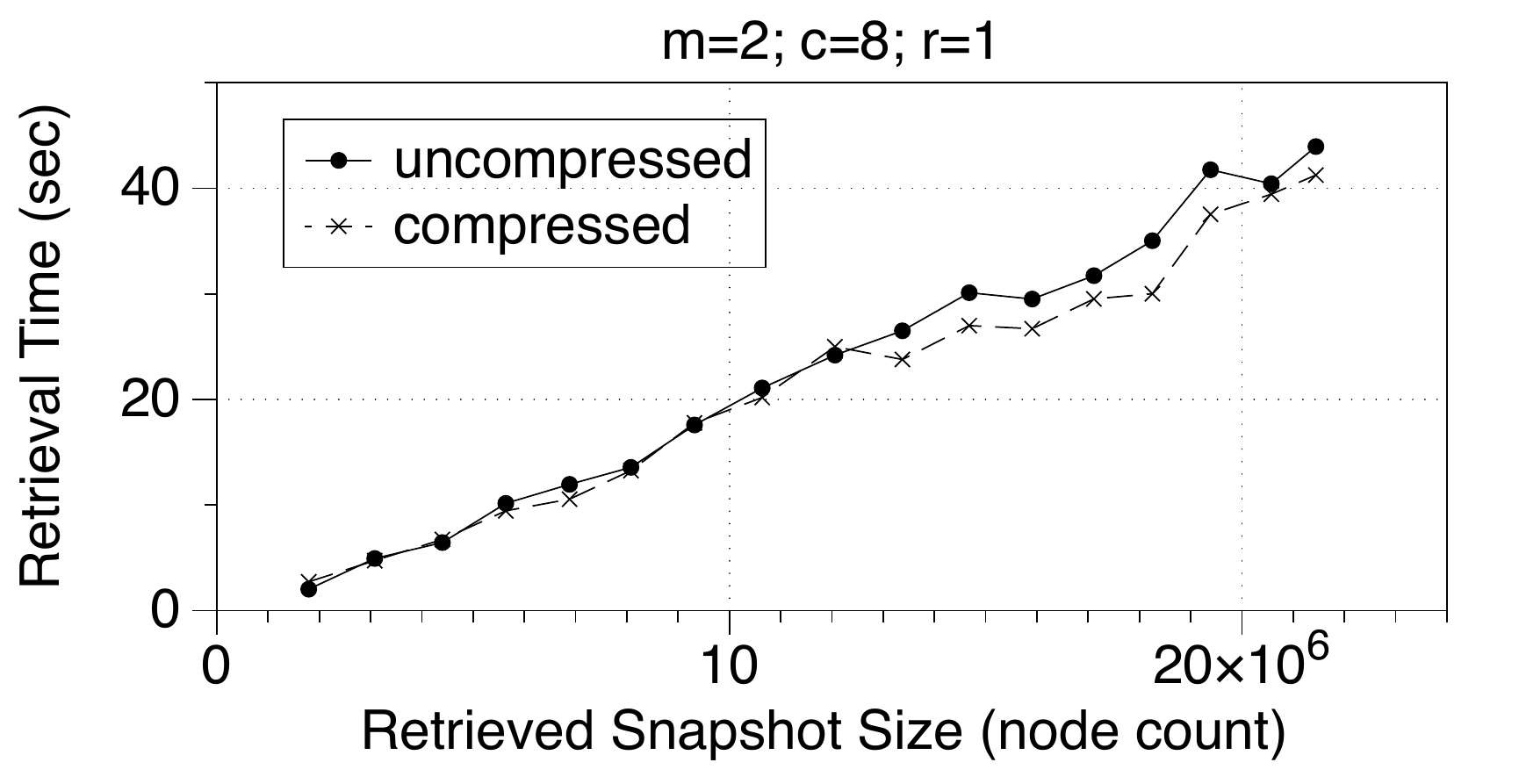}\label{fig:compression}}
\subfloat[Effect of partition sizes.]{\includegraphics[width = .33\textwidth]{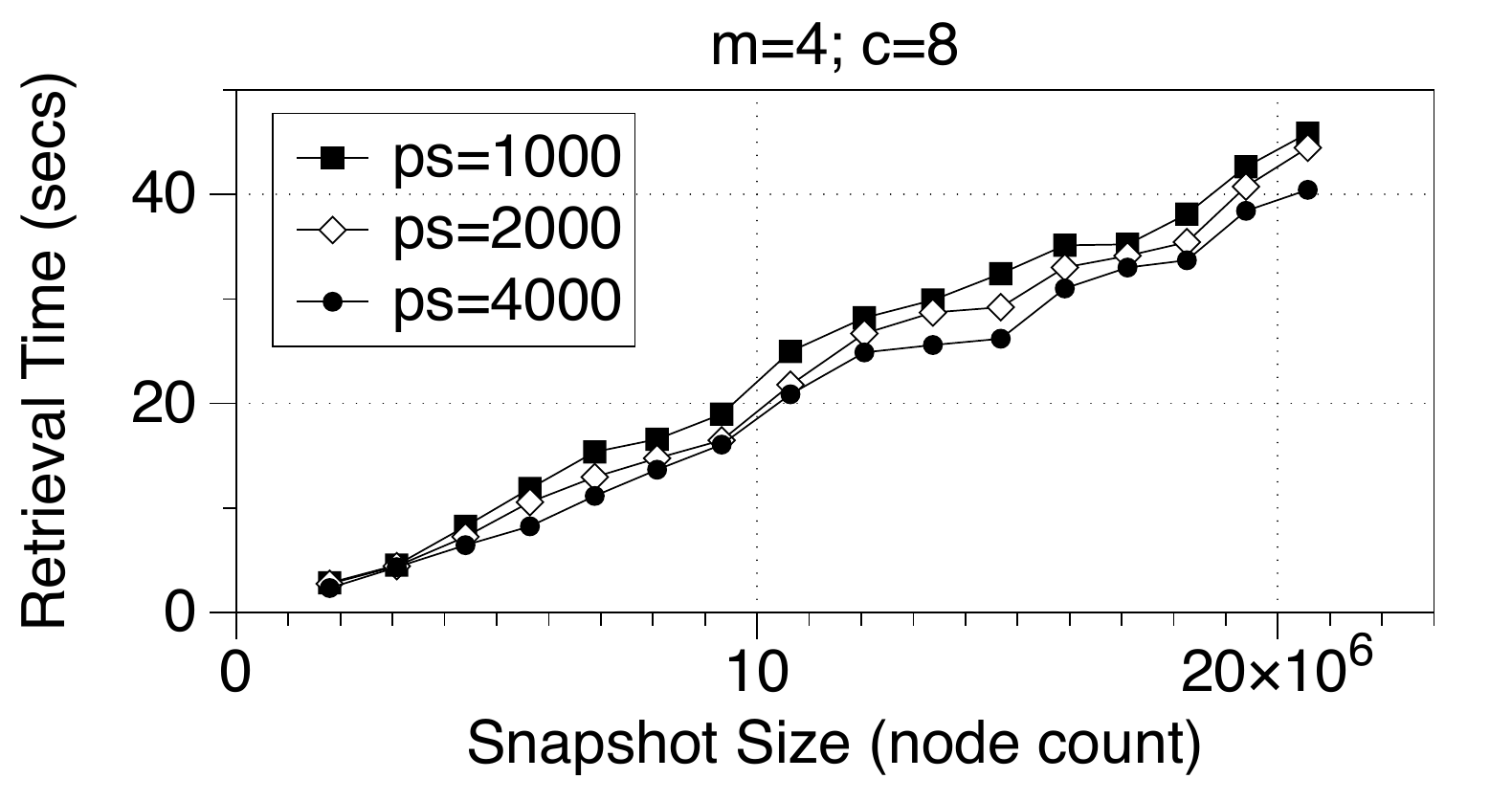}\label{fig:ss-varp}}
\subfloat[Snapshot retrieval times for Dataset 4; $m=6$; $r=1, c=1, ps=500$.]{\includegraphics[width = .33\textwidth, trim=0 0 0 20]{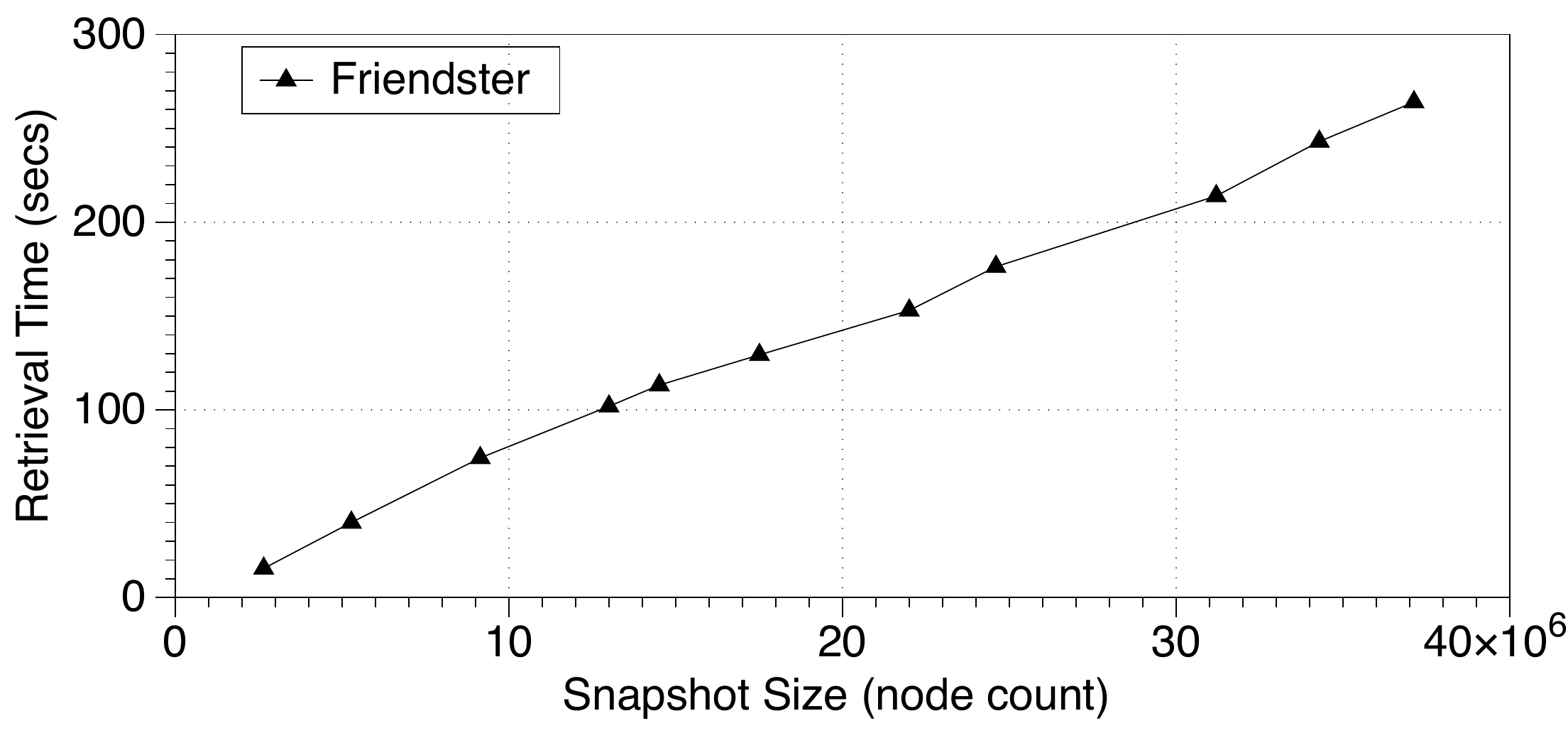}\label{fig:friendsterss}}
\caption{Snapshot retrieval across various parameters.}
\label{fig:misc-ss}
%%\vspace{-5pt}
\end{figure*}

\topic{Snapshot retrieval:} Figure~\ref{fig:ss1} shows the snapshot retrieval
times for Dataset 1 for different values of the parallel fetch factor, $c$. 
As we can see, the retrieval cost is directly proportional to the size of the output.
Further, using multiple clients to retrieve the snapshots in parallel gives 
near-linear speedup, especially with low parallelism. This demonstrates that 
TGI can exploit available parallelism well. We expect that with higher values
of $m$ (i.e., if the index were distributed across a larger number of machines), 
the linear speedup would be seen for larger values of $c$ (this is also corroborated
by the next set of experiments). The snapshot retrieval times for dataset 4 can be seen in Figure~\ref{fig:friendsterss}.

%
%\begin{figure*}
%\subfloat[Snapshot retrieval times for varying parallel fetch factor ($c$), on Dataset 1; $m=4$; $r=1, ps=500$.]{\includegraphics[width = .33\textwidth]{SS267m-MultiCore_4MachDB.pdf}\label{fig:ss1}} 
%\subfloat[Node version retrieval for Dataset 4; $m=6$; $r=1, c=1, ps=500$.]{\includegraphics[width = .33\textwidth]{NVers_Friendster_Var_c.pdf}\label{fig:friendsterNV}}
%\subfloat[Label counting in several 2-hop neighborhoods through version 
%({\tt NodeComputeTemporal}) and incremental ({\tt NodeComputeDelta}) computation, respectively. We report cumulative time taken (excluding fetch time) over varying version counts; 2 Spark workers were used for Wikipedia dataset.]{\includegraphics[width = .33\textwidth]{INCR_TAF.pdf}\label{fig:incr_exp_taf}}
%\caption{XXX}
%\label{fig:XXX}
%%\vspace{-6pt}
%\end{figure*}

%The numbers demonstrate that TGI satisfies two crucial requirements: first, the 
%snapshot retrieval cost is linearly dependent on the size of the output, and second,
%Also, multiple parallel query clients speed-up the retrieval cost as shown.

\begin{figure*}[h!]
%%\vspace{-4pt}
\subfloat[m=1; r=1; ps=500]{\includegraphics[width = .33\textwidth]{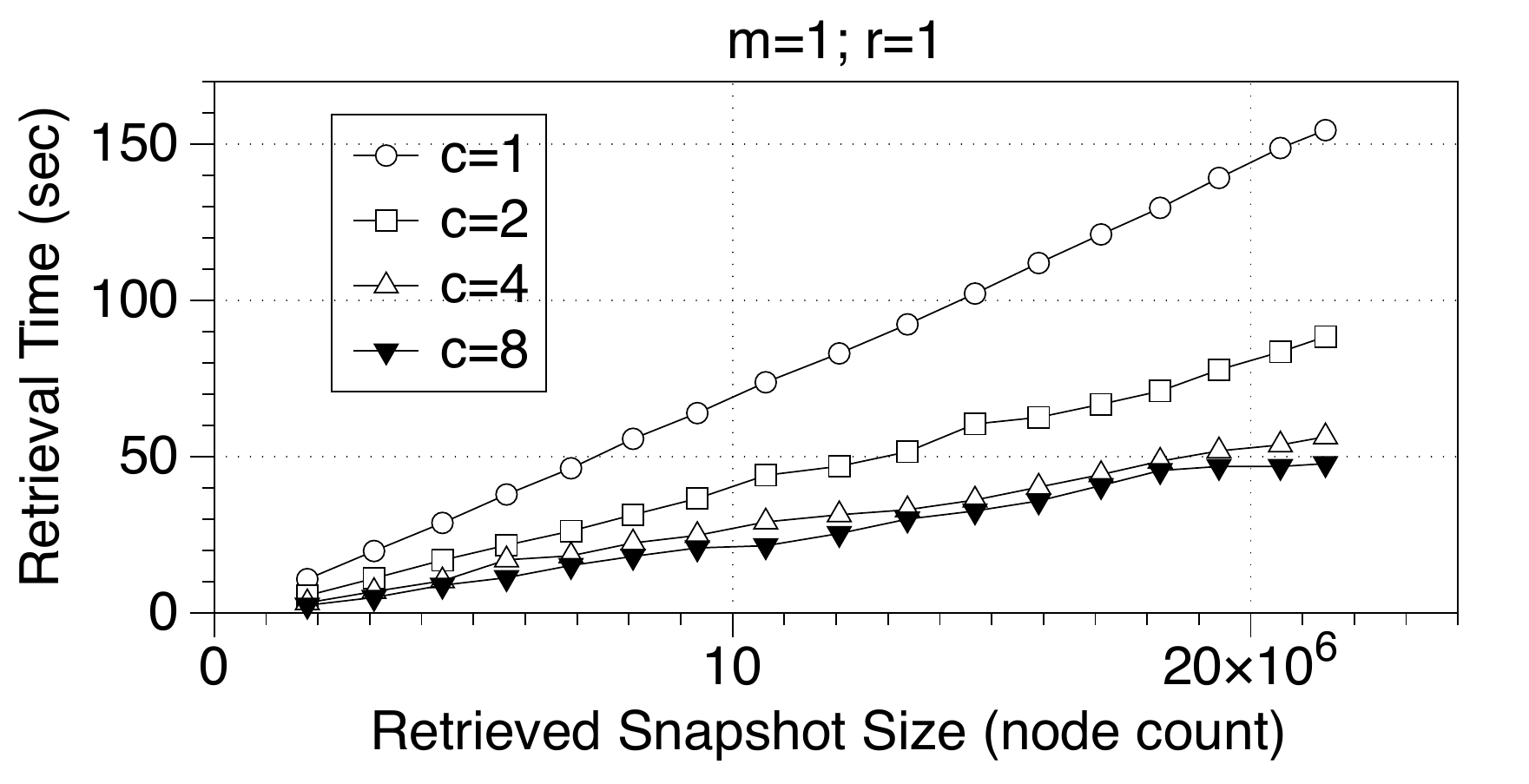}} 
\subfloat[m=2; r=1; ps=500]{\includegraphics[width = .33\textwidth]{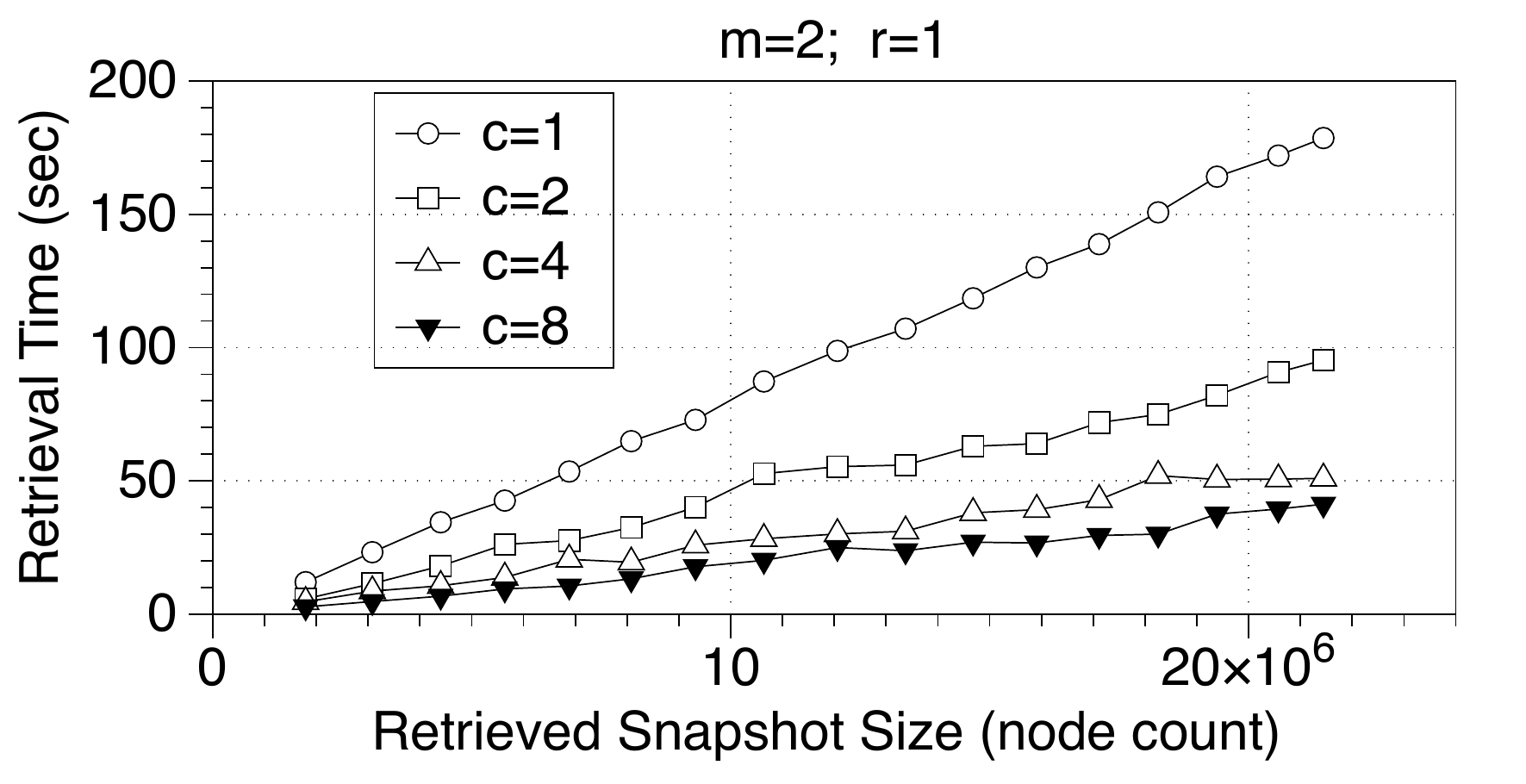}}
\subfloat[m=2; r=2; ps=500]{\includegraphics[width = .33\textwidth]{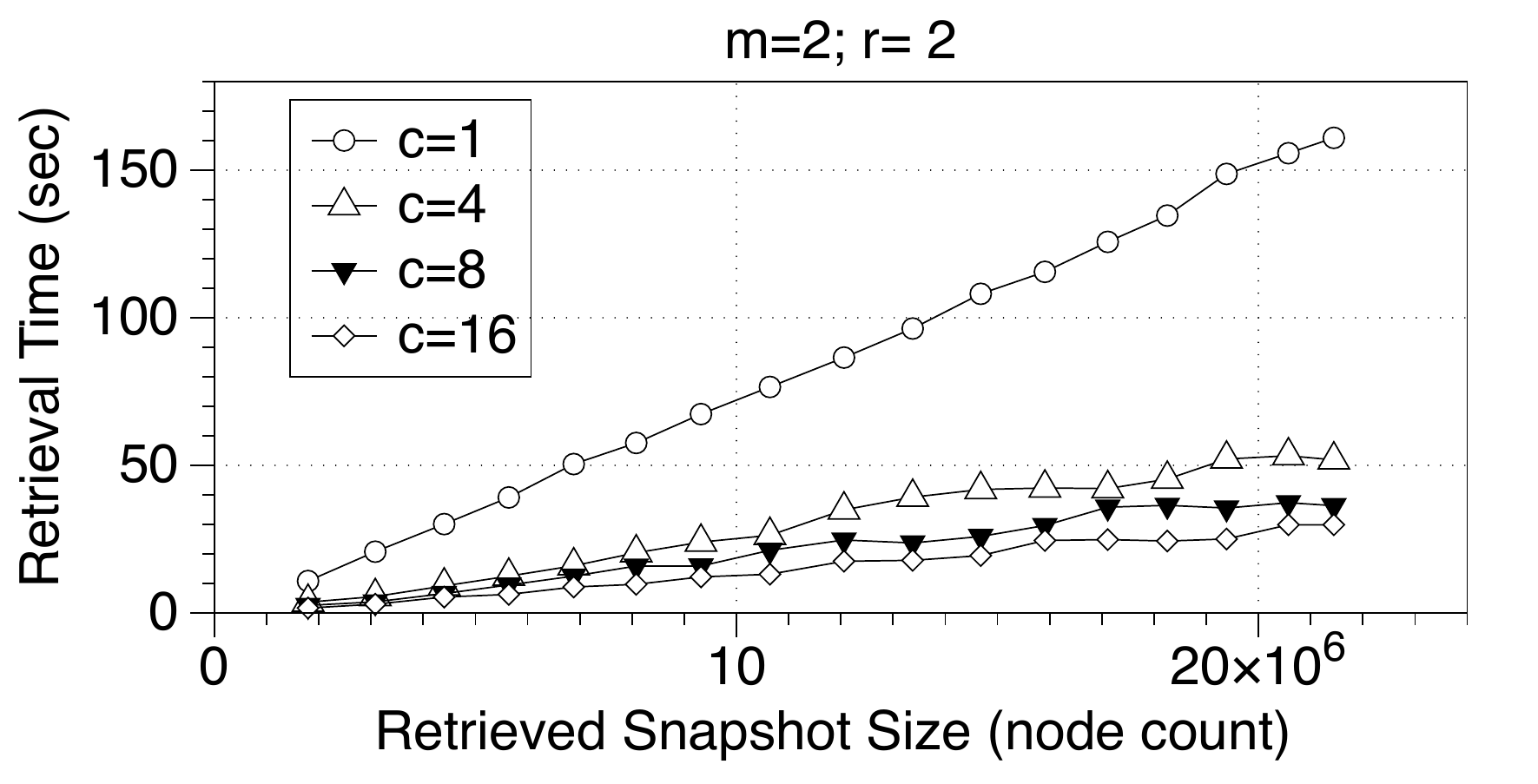}}
\caption{Snapshot retrieval times across different $m$ and $r$ values on Dataset 1.}
\label{fig:multss}
%%\vspace{-7pt}
\end{figure*}

Figure~\ref{fig:multss} shows snapshot retrieval performance for three different
sets of values for $m$ and $r$. We can see that while there is no considerable difference
in performance across the different configurations, using two storage machines
slightly decreases the query latency over using one machine, in the case of a
single query client, $c=1$. For higher $c$ values, we see that $m=2$ has a
slight edge over $m=1$. Also, the behavior for the two $m=1$ and $m=2;r=2$
cases are quite similar for same $c$ values. However, we observed that the
latter case allows a higher possibility of $c$ value whereas the former peaks
out at a lower $c$ value. %We speculate that higher $m$ and $r$ allow a higher
%throughput. 

Further, 
%Figure~\ref{fig:breakup} shows that for large
%snapshots, a bulk of the time is spent in post-processing, mostly
%{\em deserialization} of the deltas, compared to fetching the binary encoded deltas
%from the data store. 
the net effect of Cassandra compression for deltas is
negligible for TGI. We omit the detailed points of our investigation, but 
Figure~\ref{fig:compression} is representative of the general behavior. 

Size of the delta partitions (or the number) affects the performance the snapshot retrieval performance only to a small degree as seen in Figure~\ref{fig:ss-varp}.
This occurs due to a the TGI design which makes sure that all the partitions of a delta (micro-deltas) are stored contiguously in a cluster. This demonstrates that 
TGI is a superset of DeltaGraph where we are able to handle other queries along with efficient snapshot retrieval.
Note that we do not provide experimental results on the
internals of snapshot retrieval which have been thoroughly explored in
our prior work~\cite{icdepaper}. 

%XXXXX: Mention the lack of baselines and allude to the how comparison with Log approach is not necessary.

\begin{figure*}[h!]
%%\vspace{-10pt}
\subfloat[Effect of eventlist size, $l$.]{\includegraphics[width = .33\textwidth, trim = 0 5 0 10]{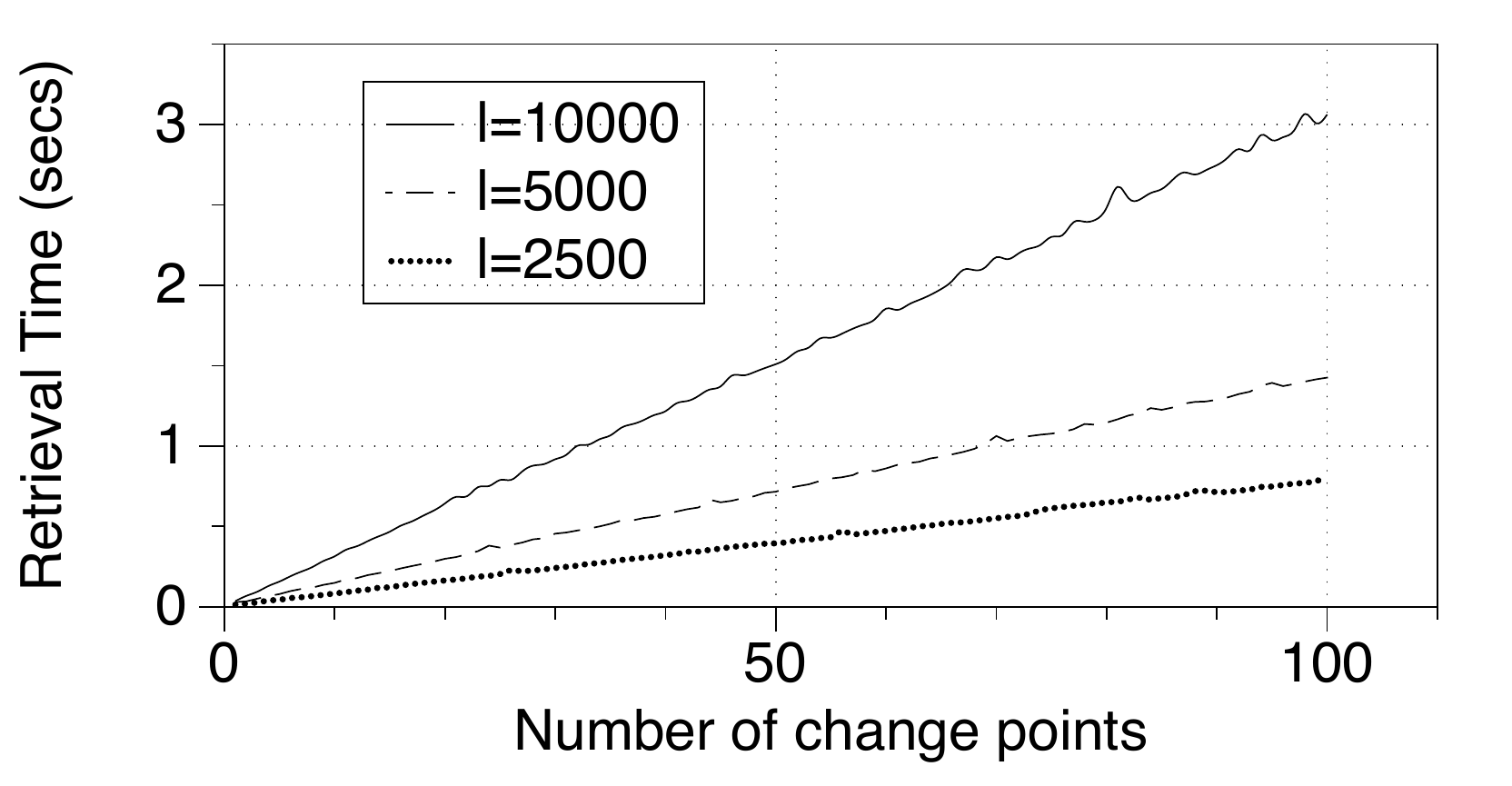}\label{fig:nver-varl}} 
\subfloat[Speedups due to parallel fetch factor, $c$.]{\includegraphics[width = .33\textwidth, trim = 0 5 0 10]{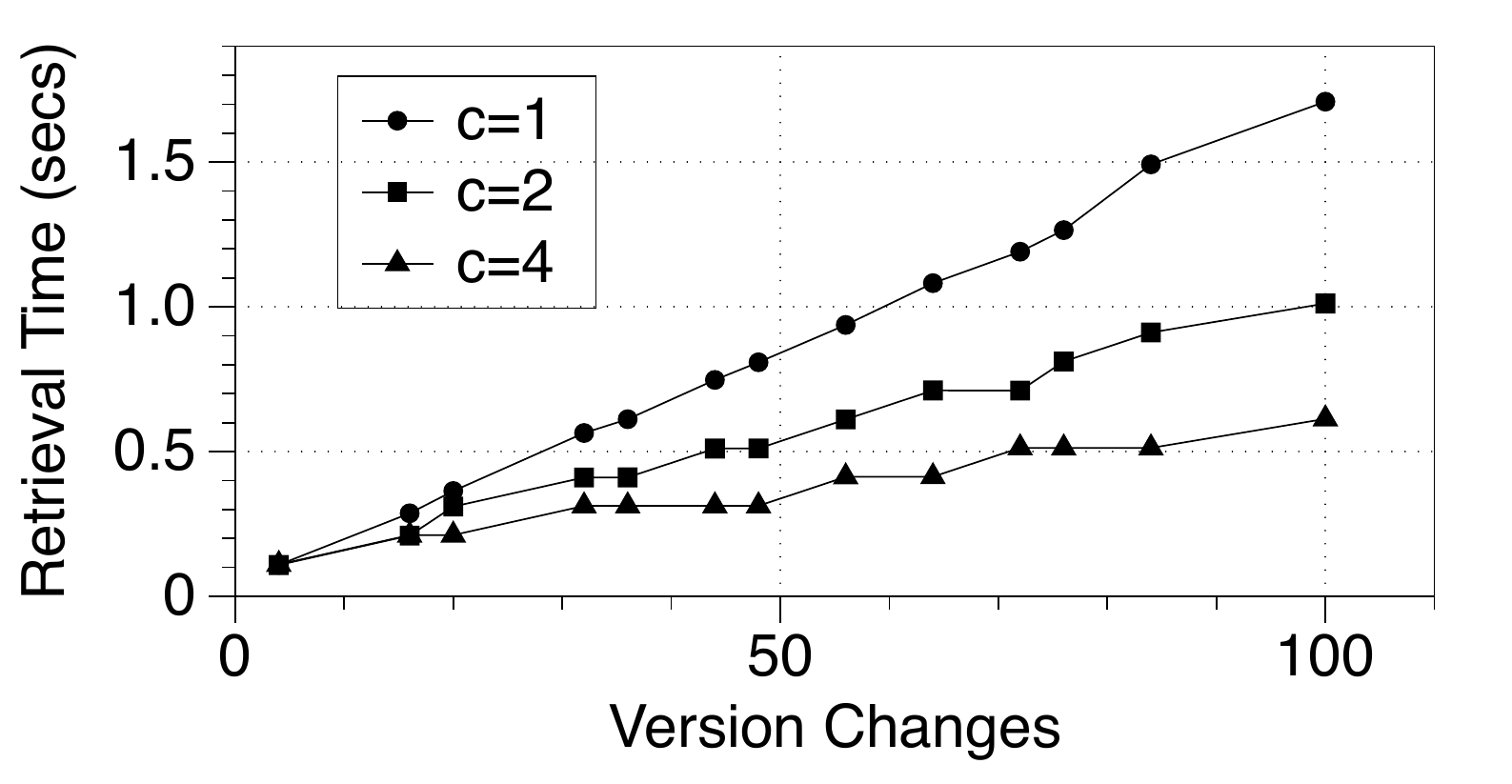}\label{fig:nver-varc}}
\subfloat[Effect of partition sizes.]{\includegraphics[width = .33\textwidth, trim = 0 5 0 10]{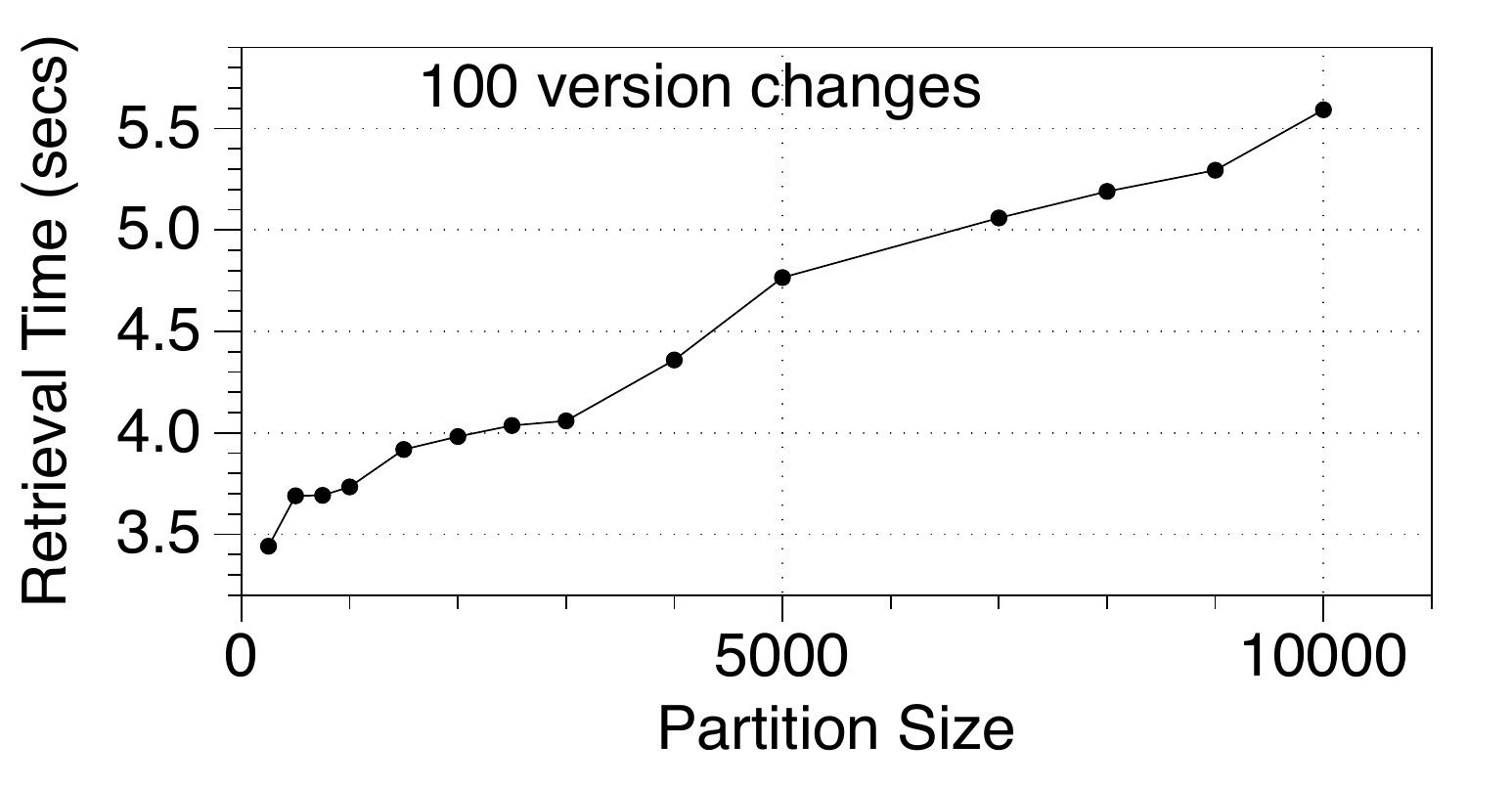}\label{fig:nver-varp}}
\caption{Node version retrieval across various parameters.}
\label{fig:misc-ss}
%%\vspace{-6pt}
\end{figure*}

\begin{figure*}[h!]
%%\vspace{-7pt}
\subfloat[][Retrieval times for 1-hop neighbor-\\hood with different partitioning and \\replication strategies.]{\includegraphics[width = .32\textwidth]{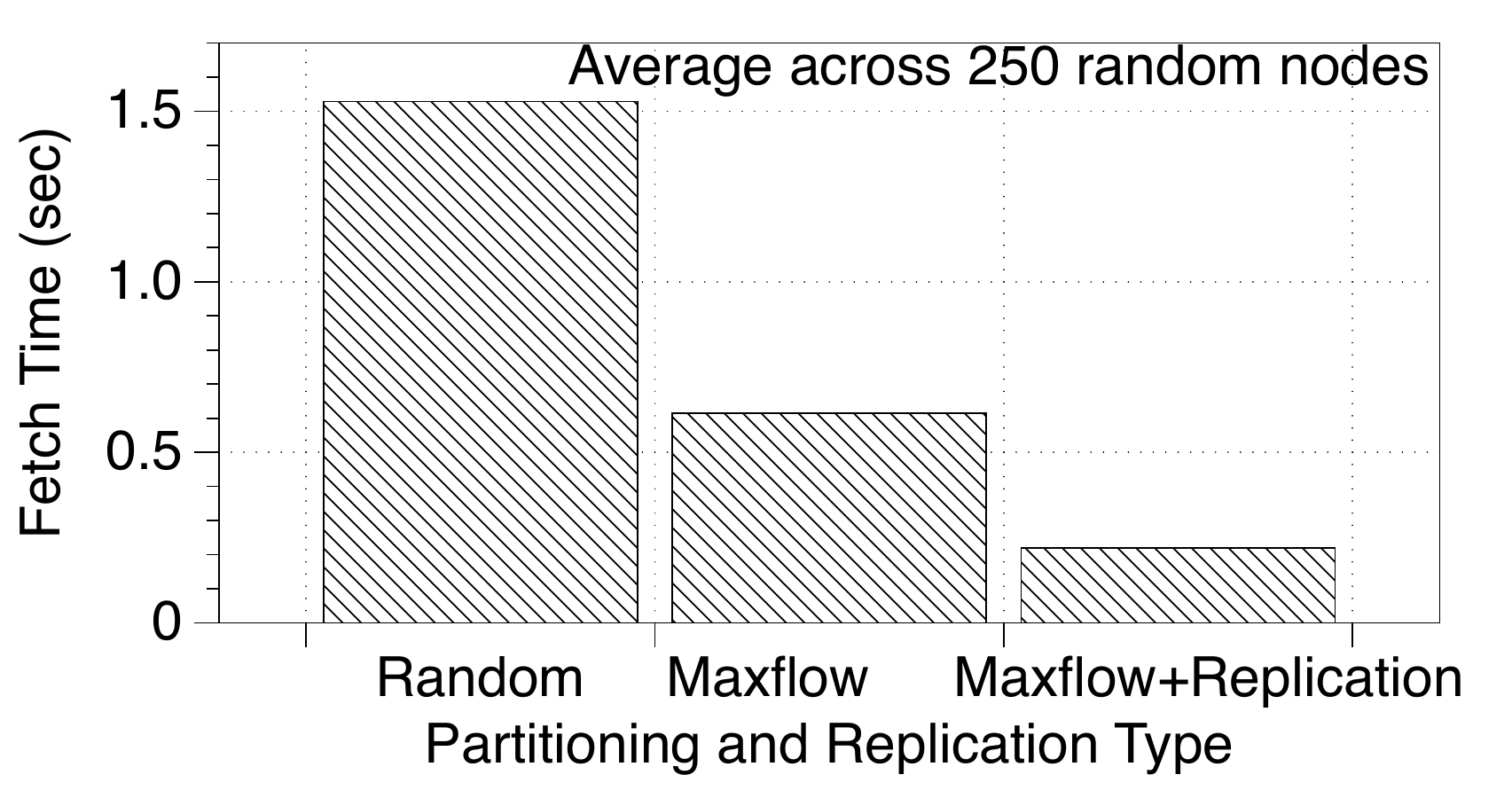}\label{fig:part_replication}} 
\subfloat[][Snapshot retrieval for varying sizes \\of datasets.]{\includegraphics[width = .33\textwidth]{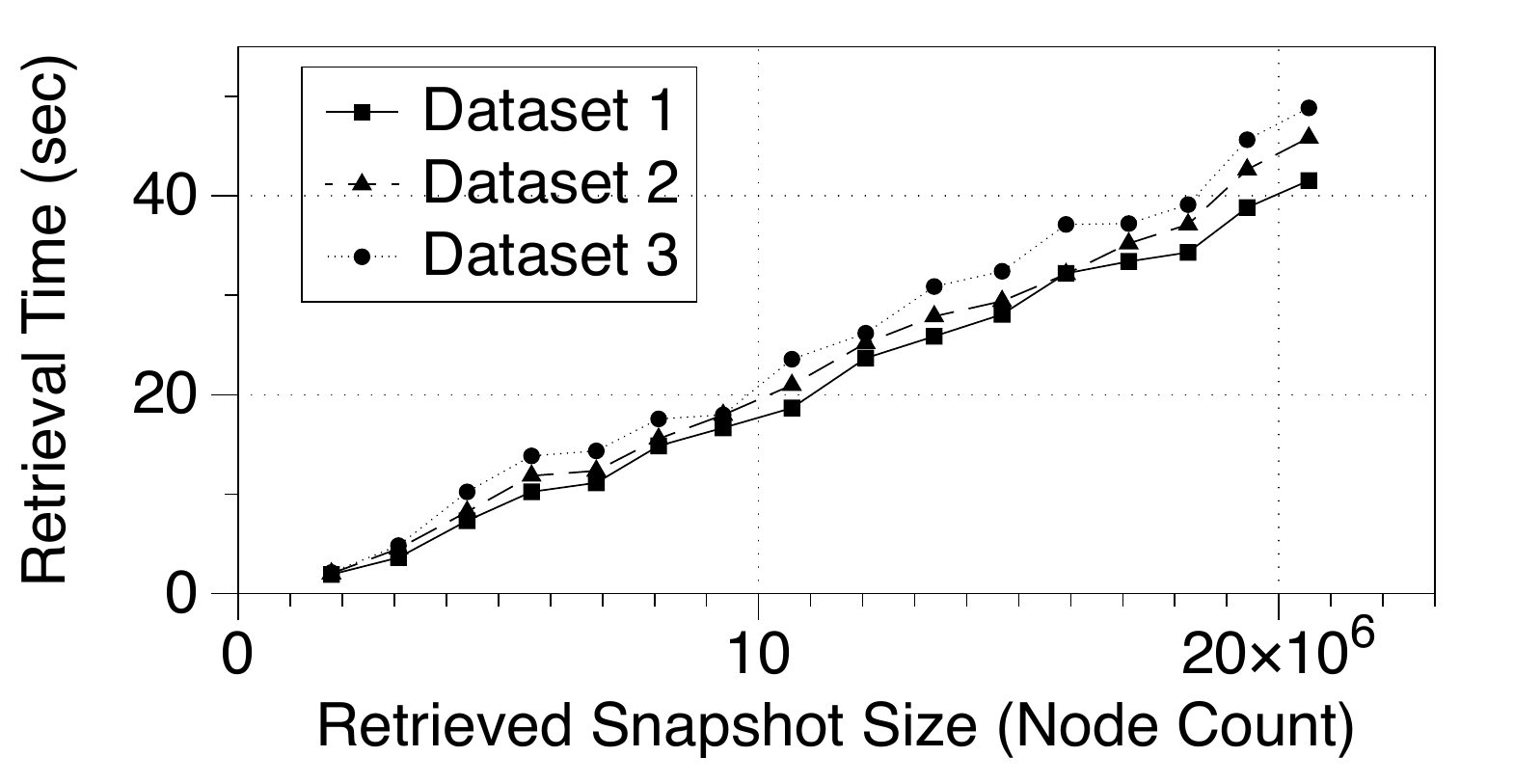}\label{fig:datascale}}
\subfloat[TAF computation times for Local Clustering Coefficient on varying graph sizes (N=node count) using different cluster sizes.]{\includegraphics[width = .33\textwidth]{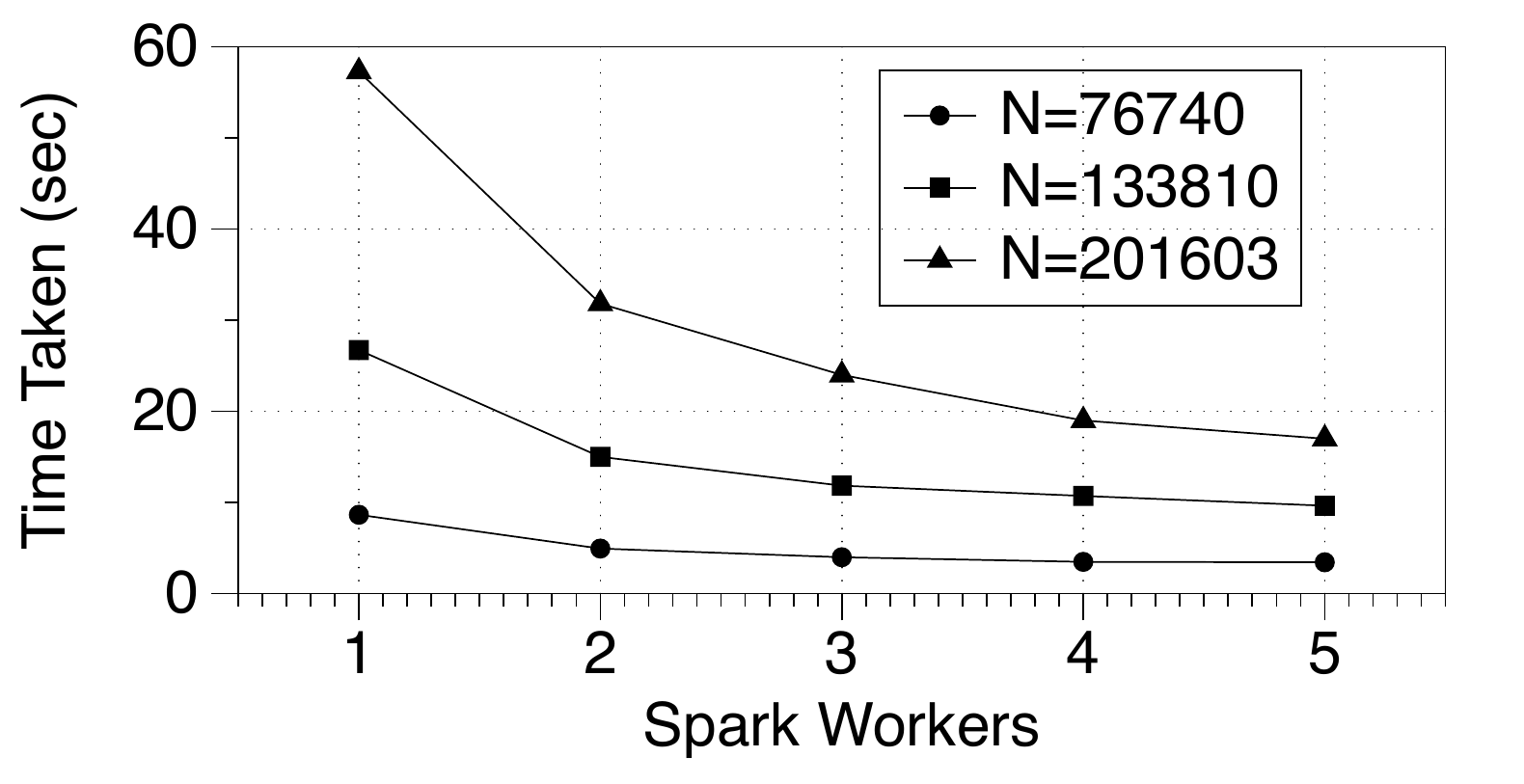}\label{fig:cc_taf}}
\caption{Experiments on partitioning type and replication; growing data size; and, TAF analytics computation.}
\label{fig:several}
%\vspace{-5pt}
\end{figure*}

\begin{figure}[h!]
\centering
\includegraphics[width=\linewidth, trim= 0 10 0 10]{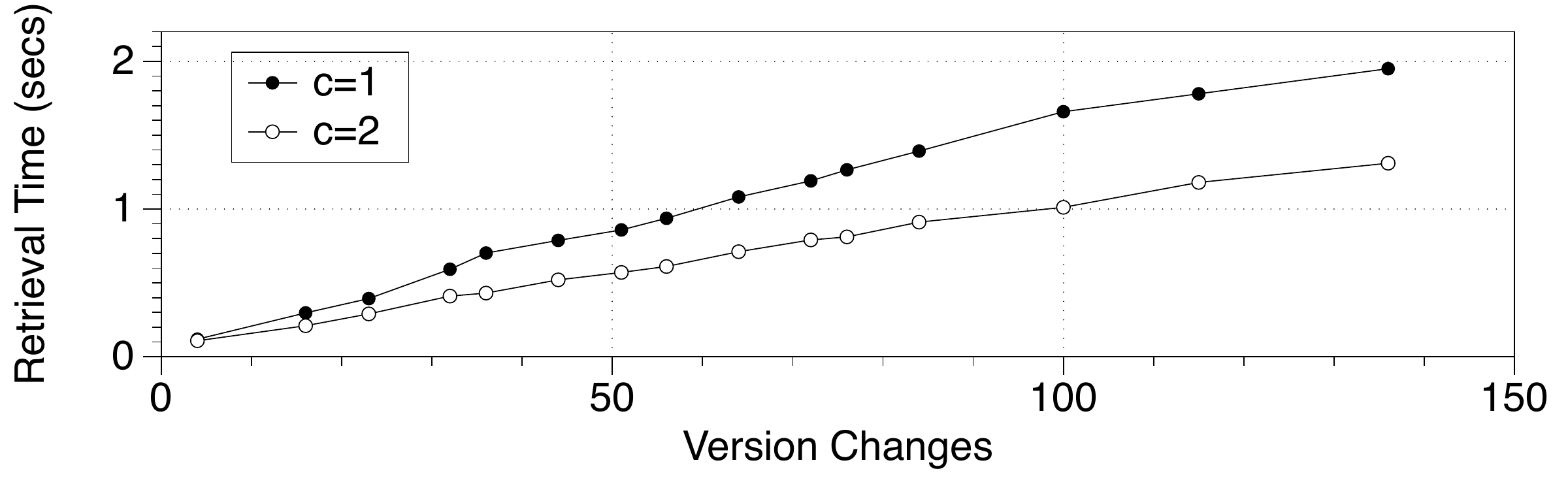}
\caption{Node version retrieval for Dataset 4; $m=6$; $r=1, c=1, ps=500$.}
\label{fig:friendsterNV}
\end{figure}

\begin{figure}
\centering
\includegraphics[width = \linewidth, trim = 0 10 0 10]{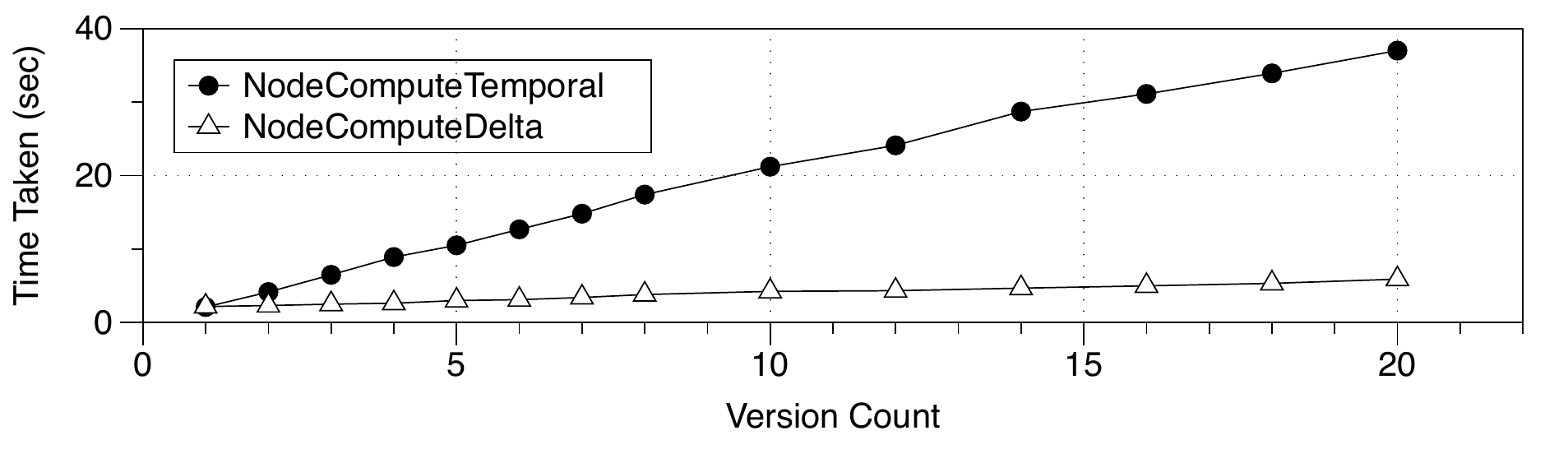} 
\caption[Label counting in several 2-hop neighborhoods through version 
(NodeComputeTemporal) and incremental (NodeComputeDelta) computation, respectively. We report cumulative time taken (excluding fetch time) over varying version counts; 2 Spark workers were used for Wikipedia dataset.]{
Label counting in several 2-hop neighborhoods through version 
({\tt NodeComputeTemporal}) and incremental ({\tt NodeComputeDelta}) computation, respectively. We report cumulative time taken (excluding fetch time) over varying version counts; 2 Spark workers were used for dataset 4.}
\label{fig:incr_exp_taf}
%%\vspace{-15pt}
\end{figure}

\topic{Node History Retrieval:} Smaller eventlists or partition sizes provide a lower latency time for retrieving different versions of a node, which can be seen in Figure~\ref{fig:nver-varl} and Figure~\ref{fig:nver-varp}, respectively. This is primarily due to the reduction in work required for fetching and deserialization. A higher parallel fetch factor is effective in reducing the latency for version retrieval (Figure~\ref{fig:nver-varc}). Note that the performance of version retrieval and snapshot retrieval with respect to varying partition sizes is contrary and represents a trade-off. However, smaller eventlist sizes benefit both version retrieval and snapshots. Node version retrieval for Dataset 4 shows a similar behavior, which can be seen in Figure~\ref{fig:friendsterNV}.

\topic{Neighborhood Retrieval:} We compared the performance of retrieving 1-hop neighborhoods, both static and specific versions, using different graph partitioning and replication choices. A topological, flow-based partitioning accesses fewer graph partitions compared to a random partitioning scheme, and a 1-hop neighborhood replication restricts the access to a single partition.This can be seen in Figure~\ref{fig:part_replication} for 1-hop neighborhood retrieval latencies. As discussed in Section~\ref{sec:tgi}, the 1-hop replication does not affect other queries involving snapshots or individual nodes, as the replicated portion is stored separately from the original partition. In case of a 2-hop neighborhood retrieval, there are similar performance benefits over random partitioning, which can be reasoned based upon similar speed-ups for 1-hop neighborhoods.

%\topic{1-hop Neighborhood Version Retrieval:}

\topic{Increasing Data Over Time:} We observed the fetch performance of TGI with an increasing size of the index. We measured the latencies for retrieving certain snapshots upon varying the time duration of the graph dataset, as shown in Figure~\ref{fig:datascale}. Datasets 2 and 3 contain additional 333 million and 733 million events over dataset 1, respectively. Only a marginal difference in snapshot retrieval performance demonstrates TGI's scalability for large datasets.

\topic{Conducting Scalable Analytics:} We examined TAF's performance through an analytical task for determining the highest local clustering coefficient in historical graph snapshot. Figure~\ref{fig:cc_taf} shows compute times for the given task on different graph sizes, as well as varying size of the Spark cluster. Speedups due to parallel execution can be observed, especially for larger datasets.

\topic{Temporal Computation:} Earlier in the chapter, we presented two separate ways of computing a quantity over changing versions of a graph (or node). Those include, evaluating the quantity on different versions of the graph separately, and alternatively, performing it in an incremental fashion, utilizing the result for the previous version and updating it with respect to the graph updates. This can be seen for a simple node label counting task in Figure~\ref{fig:computedelta_taf}. the benefits due to the incremental ({\tt NodeComputeDelta} operator) computation over a version-based computation ({\tt NodeComputeTemporal} operator) can be seen in Figure~\ref{fig:incr_exp_taf}.

\section{Conclusion}
\label{sec:conclusion}
Graph analytics are increasingly considered crucial in obtaining insights about how interconnected
entities behave, how information spreads, what are the most influential entities in the data, and many
other characteristics. %This has led to much work on graph data management systems and large-scale graph
%analysis frameworks in recent years. 
Analyzing the history of how a graph evolved can provide significant additional insights, especially about
the future. Most real-world networks however, are large and highly dynamic. This leads to creation of very
large histories, making it challenging to store, query, or analyze them. In this paper, we presented a
novel Temporal Graph Index that enables compact storage of very large historical graph traces in a 
distributed fashion, supporting a wide range of retrieval queries to access and analyze only the required 
portions of the history. We also present a distributed analytics framework, built on top of Apache Spark,
that allows analysts to quickly write complex temporal analysis tasks. Our experiments show that our temporal 
index exhibits very efficient retrieval performance across a wide range of queries, and can effectively exploit the available
parallelism in a distributed setting.
\bibliographystyle{plain}
%\nobibliography{TGIdraft}
\bibliography{TGIdraft-extended}

\end{document}